\documentclass[twocolumn]{aastex63}
\usepackage{amsmath}
\DeclareMathOperator*{\argmax}{arg\,max}

\newcommand{\Rcld}{R_{\rm cl}}
\newcommand{\Mcld}{M_{\rm cl}}
\newcommand{\Scld}{\Sigma_{\rm cl}}
\newcommand{\SEdd}{\Sigma_{\rm Edd}}
\newcommand{\Savg}{\bar\Sigma }
\newcommand{\Msun}{M_{\odot}}
\newcommand{\Mst}{M_*}
\newcommand{\Mstsnk}{M_{*,{\rm sink}}}
\newcommand{\Mw}{M_w}
\newcommand{\Mgas}{M_g}
\newcommand{\Mclw}{M_{{\rm cloud},w}}
\newcommand{\dMclw}{\dot{M}_{{\rm cloud,ej}}}
\newcommand{\Mcut}{M_{\rm cut}}
\newcommand{\Lbox}{L_{\rm box}}
\newcommand{\tbr}{t_{\rm br}}
\newcommand{\fwgrd}{f_{w,{\rm grid}}}
\newcommand{\fwind}{f_{\rm wind}}
\newcommand{\fwnds}{f_{\rm wind},*}
\newcommand{\presc}{p_{r,{\rm esc}}}
\newcommand{\pc}{\, {\rm pc}}

\newcommand{\Lwind}{\mathcal{L}_w}
\newcommand{\pdotw}{\dot{p}_w}
\newcommand{\Vw}{V_w}
\newcommand{\reff}{\mathcal{R}_b}
\newcommand{\Ecool}{\dot{E}_{\rm cool}}
\newcommand{\Eleak}{\dot{E}_{\rm leak}}
\newcommand{\eof}{\varepsilon_{\rm of}}
\newcommand{\sfe}{\varepsilon_*}
\newcommand{\sfef}{\varepsilon_{*,{\rm f}}}
\newcommand{\sfeff}{\varepsilon_{\rm ff}}
\newcommand{\tffo}{t_{\rm ff,0}}
\newcommand{\tff}{t_{\rm ff}}
\newcommand{\rhobar}{\overline{\rho}}
\newcommand{\slnS}{\sigma_{\ln \Sigma}}

\received{XXX}
\revised{XXX}
\accepted{\today}
\submitjournal{ApJL}

\shorttitle{Stellar Wind Regulated Star Formation and Self-Pollution}
\shortauthors{Lancaster et al.}
\graphicspath{{./}{figures/}}
\begin{document}

\title{Star Formation Regulation and Self-Pollution by Stellar Wind Feedback}

\correspondingauthor{Lachlan Lancaster}
\email{lachlanl@princeton.edu}

\author[0000-0002-0041-4356]{Lachlan Lancaster}
\affiliation{Department of Astrophysical Sciences, Princeton University, 4 Ivy Lane, 08544, Princeton, NJ, USA}

\author[0000-0002-0509-9113]{Eve C. Ostriker}
\affiliation{Department of Astrophysical Sciences, Princeton University, 4 Ivy Lane, 08544, Princeton, NJ, USA}

\author[0000-0001-6228-8634]{Jeong-Gyu Kim}
\affiliation{Department of Astrophysical Sciences, Princeton University, 4 Ivy Lane, 08544, Princeton, NJ, USA}

\author[0000-0003-2896-3725]{Chang-Goo Kim}
\affiliation{Department of Astrophysical Sciences, Princeton University, 4 Ivy Lane, 08544, Princeton, NJ, USA}

\begin{abstract}
Stellar winds contain enough energy to easily disrupt the parent cloud surrounding a nascent star cluster, and for this reason have been considered candidates for regulating star formation. However, direct observations suggest most wind power is lost, and \citet{Lancaster21a,Lancaster21b} recently proposed that this is due to efficient mixing and cooling processes. Here, we simulate star formation with wind feedback in turbulent, self-gravitating clouds, extending our previous work. Our simulations cover clouds with initial surface density $10^2-10^4\, M_\odot\ \pc^{-2}$, and show that star formation and residual gas dispersal is complete within 2 - 8 initial cloud free-fall times.  The ``Efficiently Cooled'' model for stellar wind bubble evolution predicts enough energy is lost for the bubbles to become momentum-driven, we find this is satisfied in our simulations. We also find that wind energy losses from turbulent, radiative mixing layers dominate losses by ``cloud leakage'' over the timescales relevant for star formation.  We show that the net star formation efficiency (SFE) in our simulations can be explained by theories that apply wind momentum to disperse cloud gas, allowing for highly inhomogeneous internal cloud structure. For very dense clouds, the SFE is similar to those observed in extreme star-forming environments.  Finally, we find that, while self-pollution by wind material is insignificant in cloud conditions with moderate density (only $\lesssim 10^{-4}$ of the stellar mass originated in winds), our simulations with conditions more typical of a super star cluster have star particles that form with as much as 1\% of their mass in wind material.
\end{abstract}


\section{Introduction}
\label{sec:intro}

Self-gravitating molecular clouds are observed to be star-forming in a wide range of environments, from dwarf galaxies \citep{Lopez11,Cheng21}, to the Solar neighborhood \citep{SOFIAFeedback20,FujitaOrion21}, to galactic center starbursts \citep{Johnson15,Leroy18,Levy20,Emig20}. On the scale of molecular clouds, the lifetime star formation efficiency (SFE) depends on the processes that truncate star formation (SF) by destroying the cloud, some of which are external and some of which are internal \citep[e.g.][]{ChevanceRev20}. For giant molecular clouds (GMCs) that create massive stars, internal processes  associated with SF feedback dominate the regulation of this process. There are many mechanisms by which these massive stars act to disperse their surrounding clouds, and it is likely that the mechanism that plays the largest role depends on the physical conditions of the surrounding cloud \citep[e.g.][]{KMBH19}.

Previous numerical work has extensively  investigated the effects of radiation feedback from star clusters on the surrounding GMC -- i.e. ionized gas pressure, direct radiation pressure, and reprocessed radiation pressure, for a range of cloud conditions \citep[e.g.][]{Dale12,Dale_2013a,Skinner_Ostriker2015,Raskutti16,Raskutti17,Howard17,JGK18,JGK19,KOF_2021}. In these models, photoionization usually dominates cloud destruction for ``normal'' GMCs \textbf{($\Scld \sim 10^2 \, \Msun\, {\rm pc}^{-2}$)}, while radiation pressure becomes relatively more important for clouds with very high surface density, following general trends predicted with analytic models \citep[e.g.][]{Krumholz_Matzner2009,Kim_JG2016}. The direct injection of kinetic energy and mass (so-called ``mechanical'' feedback) may also be important -- sources are stellar winds at early stages followed by supernovae (SNe) \citep{Agertz_2013,Rogers_Pittard2013,Walch_Naab2015,Iffrig_Hennebelle2015,Geen_2016}. For clouds that are relatively diffuse and have long evolutionary timescales, SNe are effective, and indeed SFRs in large-scale simulations with SN feedback are consistent with observations for these conditions even when dynamical effects of radiation are omitted \citep[e.g.][]{CGK_TIGRESS1}. For very high surface density clouds, where super star clusters (SSCs) form, the SF timescales are too short for SNe to affect evolution significantly, but winds can potentially play a key role.

In \citet{Lancaster21a} we proposed a new theoretical model for the expansion of stellar wind bubbles under conditions present in the turbulent molecular clouds where stars are born. This model appeals to efficient loss of energy through turbulent, fractal, radiative mixing layers between the hot, shocked wind and the surrounding cloud. This results in bubble evolution that is momentum-driven ($\dot p_b = const.$, equivalent bubble radius $R_b \propto t^{1/2}$, 
also as in \citet{Steigman75}) rather than energy-driven ($\dot E_b= const.$, $R_b \propto t^{3/5}$) as in the classical \citet{Weaver77} solution.  In \citet{Lancaster21b}, we presented a large suite of three-dimensional (3D) hydrodynamical simulations to validate this theory.  However, our previous work did not take into account the effects of gravity, either for dynamical evolution of the large-scale cloud, or for establishing a self-consistent star formation history (SFH) by inducing small-scale collapse. 

Investigating the interplay between gravity, turbulence, and stellar wind feedback is the main goal of the present work. A key motivation is to confirm that energy losses are still large, consistent with the \citet{Lancaster21a} solution. If not, it would be difficult to understand the observed high SFEs of SSC-forming clouds \citep{Johnson15,Leroy18,Emig20,Levy20}. If evolution were to follow the wind-driven bubble solution from \citet{Weaver77}, the large momentum injection from an inefficiently-cooling bubble would easily destroy GMCs even with very low SFE and very high surface density \citep{Lancaster21a}. We have previously argued that with much-reduced momentum injection driven by cooling, winds would be compatible with observed high SFEs in SSC-forming clouds; here we directly test this prediction with simulations. 
%\textbf{Past work has also appealed to so-called `catastrophic' cooling of energy out of winds (although through a different physical mechanism than \citet{Lancaster21a}) to help explain how wind material can form a second generation of stars in globular clusters \citep{Wunsch17,LochhaasThompson17,TT19}.}

In real clouds, multiple feedback processes operate simultaneously, and several numerical simulations have included treatment of both wind and radiation feedback in some form \citep{Dale14,Geen21,Wall20,Grudic21}.  Here, our goal is instead to isolate the effects of winds independent of other feedback processes, and to systematically explore cloud dynamical evolution and SFH for varying parent cloud properties.  This allows us to develop and test simple theoretical models, while providing a baseline for more comprehensive studies.   

Our self-gravitating models allow us to investigate another important question. In the literature, two modes of energy loss have been proposed in order to explain weak observed signatures of winds: enhanced radiative cooling \citep{Rosen14,Lancaster21a} \textit{or} bulk advection of wind energy out of the cloud -- so called ``leakage'' \citep{HCM09}. In \citet{Lancaster21b}, we showed that interface mixing leading to strong cooling can be very effective, and explains the low pressure of X-ray emitting gas observed in pre-SN systems \citep[see discussion in][]{Lancaster21a}. Here we are able to investigate the roles of mixing/cooling vs. leakage in removing energy from winds that develop in a cloud with self-consistent star formation. As we shall show, both are important but at different times.

Beyond their potential importance in helping to control star formation, winds are also interesting for the way in which they act to chemically pollute the gas in a molecular cloud, leading to so-called ``self-enrichment'' of subsequent populations of stars that form within the molecular clouds. The details of this process are important to fields ranging from galactic archaeology, where the process of self-enrichment may undermine the strategy of so-called ``chemical tagging'' \citep{FBH02}, to multiple stellar populations with different abundances within globular clusters (GCs), where nearly all proposed models for explaining these populations involve ``self-enrichment'' \citep{Wunsch17,LochhaasThompson17,Bailin18,TT19,BastianLardo18}. To be clear, the winds from massive OB stars explored here do not have drastically different chemical abundances from progenitor gas, especially before the Wolf-Rayet phase, and therefore would not, themselves, cause strong self-enrichment. In tracking the re-integration of wind material in subsequent stars we provide a first look at how this process may work for stellar winds more generally. For this reason we refer to the process explored here as ``self-pollution.''

The plan of this {\it Letter} is as follows. In \autoref{sec:methods} we briefly review our numerical methods, explained in full elsewhere. Results of our simulations, and connection to other theoretical work, is presented in \autoref{sec:results}. Finally, we summarize our conclusions and connect to observations in \autoref{sec:conclusions}.

\section{Methods}
\label{sec:methods}

We simulate the collapse and subsequent dispersal by stellar wind feedback of self-gravitating, turbulent gaseous clouds. The simulation framework we employ is built on the 3D magneto-hydrodynamics (MHD) code \textit{Athena} \citep{Stone08_Athena}, though we do not include magnetic fields in this study. We run all of our simulations on a fixed, Cartesian grid using the Roe approximate Riemann solver \citep{Roe81}, second order spatial reconstruction, and the unsplit van Leer integrator \citep{StoneGardiner09}.

We include cooling in the gas as implemented in the \textit{Athena}-TIGRESS code base \citep{CGK_TIGRESS1}. This cooling module assumes a simple cooling coefficient $\Lambda(T)$ using the fitting formula of \citet[][see \citet{Kim08} for corrections]{KoyamaInutsuka02} for $T< 10^{4.2}\, {\rm K}$ gas and the tabulated results from \cite{SutherlandDopita93} for $T>10^{4.2}\, {\rm K}$ gas (assuming collisional ionization equilibrium).  We adopt a uniform background heating rate of $2\times 10^{-26}\, {\rm erg}\, {\rm s}^{-1} \, {\rm H}^{-1}$ which decreases smoothly to zero in higher temperature gas.

We employ sink/star particles as implemented in \citet{GongOstriker13}, with modifications described in \citet{Kim_Ostriker_SMAUG2020}. Sink particles, representing stellar (sub-) clusters, are created when a density threshold  \cite[based on the collapse solution of][]{Larson69,Penston69} is reached, the cell forming the particle is at a minimum of the gravitational potential, and the flow is converging along all grid directions.

\begin{figure*}[t!]
    \centering
    \includegraphics[width=\textwidth]{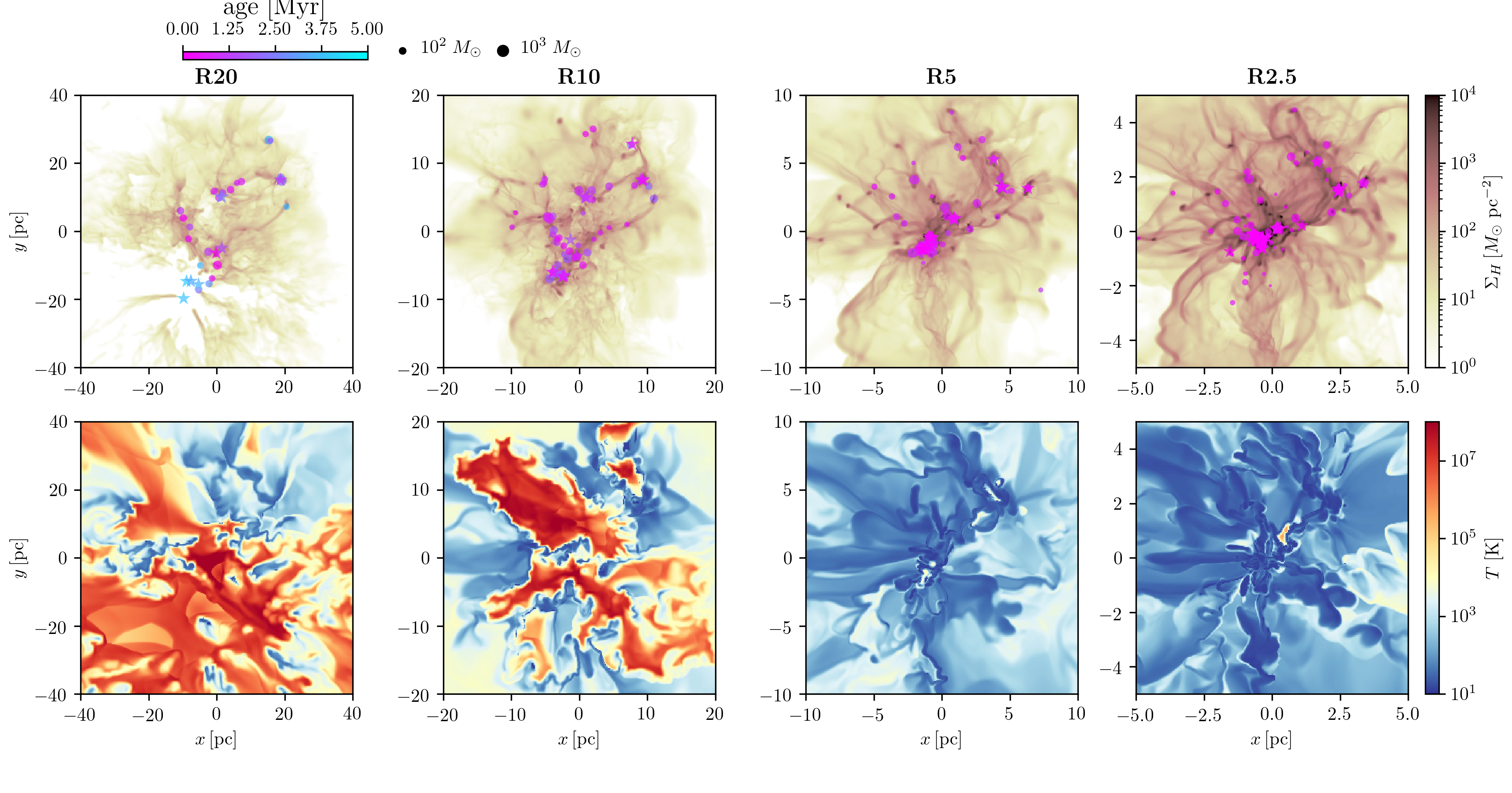}
    \caption{Snapshots from simulations at the time when 75\% of the final stellar mass has formed. Models {\bf R20} to {\bf R2.5} are shown from left to right, with all simulations shown here having the same initial turbulence realization (with differing amplitudes). {\it Top row}: the integrated gas surface density, together with projected locations of star particles colored by age with size scaled by mass. The ``active'' star particles producing winds are shown as stars while inactive star particles are shown as circles. {\it Bottom row}: slices of gas temperature (centered on the center of mass of the star particles).}
    \label{fig:snapshot}
\end{figure*}

We include star formation feedback in the form of energy and momentum representing the collective stellar wind inputs from a coeval population of O and B stars \citep{Vink01,Vink21,Leitherer92}. We use prescriptions for the energy and momentum from Starburst99 \citep{SB99,Leitherer14} for two different initial stellar mass functions (IMFs, described further below), both employing the Geneva, non-rotating stellar evolutionary tracks \citep{Ekstrom12}. These calculations use the wind prescriptions of \cite{Vink01}, which are thought to over-estimate the mass-loss rates, $\dot{M}_w$, by a factor of 2-3 \citep{Smith14}; $\pdotw = \sqrt{2\Lwind\Mw}$ would also be affected by this. This energy and momentum is injected within a radius $r_{\rm fb}/\Delta x = 3.2$ of the star particle using the hybrid kinetic/thermal injection scheme described in \cite{Lancaster21b}, which employs a subcell method for dividing the energy and (vector) momentum more accurately over the injection region \citep[see also][]{Ressler20}. When mass from a wind is injected to the grid, we also inject an equal mass in a passive scalar which is advected with the grid gas, allowing us to track subsequent mixing of the wind material.

In our simulations, we impose a minimum mass $\Mcut = 10^3 \Msun$ for the initialization of a stellar wind from a sink particle. That is, for all sink particles with mass $\Mstsnk < \Mcut$ accretion is permitted, but for higher mass particles accretion is turned off and wind injection proportional to $\Mstsnk$ is initiated. We do this both to reduce the computational cost of injecting wind material and to represent the correlation of winds with high mass stars, which are relatively rare.  In addition, since the most massive stars contribute a disproportionate fraction of the power, the number of ``important'' wind sources is limited. The mean specific wind luminosity and mass loss rates increase while the variances decrease for more fully-sampled mass functions. For example, clusters of mass $500,\, 1000,$ and $5000\, \Msun$ sampled from a Kroupa IMF have respective expectation values of the wind luminosity per stellar mass ${\cal L}_w/\Mst = 4\times 10^{32}$, $10^{33}$, and $4\times 10^{33}\, {\rm erg/s/}M_{\odot}$ with inter-quartile ranges spanning $1.8$, $1.5$, and $0.5$ dex, respectively. All of our simulations have total mass $>10^4 \Msun$, well into the  fully-sampled regime, so it is appropriate to assign wind power based on a fully sampled IMF.

If we make $\Mcut$ too low, the simulations become expensive due to having many sources, and they are also unphysical in having the wind power nearly equally divided among an unrealistically large number of sources.  If we make $\Mcut$ too high, we artificially suppress feedback. We therefore choose $\Mcut = 10^3 \Msun$ as a compromise. For a cluster of mass $3 - 9\times 10^4 \, \Msun$ (the range for our simulations) divided among particles of mass $10^3 \Msun$, there would be 30-90 active sources.  This can be compared to the expectation value 29-85 for the number of stars that would provide $95 \%$ of the wind luminosity in a cluster of this mass range that fully samples the Kroupa IMF\footnote{Calculations in the last two paragraphs were performed using the \cite{Leitherer92} wind prescriptions along with the \cite{MIST} isochrones at solar metallicity, without rotation, resulting in small differences between the results of \cite{Leitherer92}.}.

We explore the implications of our prescription and parameter choice, and compare with alternative ways of assigning feedback to star particles, in \autoref{app:mcut}.  It is important to note that our simulations are not suited to studying the mass spectrum of either stars or clusters; the initial mass spectrum of sink particles depends on grid resolution, and their growth via accretion is truncated when winds turn on after $M=\Mcut$.

\begin{deluxetable*}{cccccccccccc}
\tablecaption{Parameters and results of simulation suite.\label{tab:sim_params}}
\tablewidth{0pt}
\tablehead{Model name & Cloud radius & \colhead{$\overline{n}_{\rm H, cl}$}  & $v_t$ & $t_{{\rm ff}, 0}$ & $\Delta x$%\tablenotemark{a}
& $\sfef$ & $\varepsilon_{*, {\rm nfb}}$ & $\tbr$ & $t_{\rm dest}$ & $t_{\rm SF}$ & $t_{\rm pr}$ \\
& $[{\rm pc}]$ & \colhead{$[{\rm cm}^{-3}]$} & $[{\rm km}\, {\rm s}^{-1}]$  & $[{\rm Myr}]$ & $[{\rm pc}]$ & \% & \% & $[{\rm Myr}]$ & $[{\rm Myr}]$ & $[{\rm Myr}]$ & $[{\rm Myr}]$}
%\decimalcolnumbers
\startdata
{\bf R20}  & 20  & 86.3  & 5.08 & 4.68 & $0.31$ & $28^{+2}_{-7}$ & $75^{+5}_{-4}$ & $8.1 \pm 1.4$       & $14.8^{+5.4}_{-0.6}$ & $8.2^{+2.8}_{-1.2}$  & $6.1^{+0.4}_{-0.2}$ \\
{\bf R10}  & 10  & 690  & 7.18 & 1.66 & $0.16$ & $51^{+4}_{-10}$ & $82^{+3}_{-1}$ & $5.0^{+0.6}_{-1.0}$ & $6.1^{+0.7}_{-0.5}$  & $3.6^{+0.8}_{-0.5}$  & $4.3^{+0.2}_{-0.7}$ \\
{\bf R5}   & 5   & 5520  & 10.2 & 0.586 & $0.08$ & $71^{+3}_{-4}$& $87\pm 1$ & $3.1 \pm 0.5$       & $3.5^{+0.3}_{-0.4}$  & $2.1^{+0.1}_{-0.3}$  & $2.9^{+0.2}_{-0.4}$ \\
{\bf R2.5} & 2.5 & 44200 & 14.4 & 0.207 & $0.04$ & $85^{+1}_{-3}$ & $88\pm 1$ & $1.7^{+0.2}_{-0.1}$ & $1.8^{+0.04}_{-0.2}$ & $1.0^{+0.2}_{-0.03}$ & $1.8\pm 0.1$       \\
\enddata
\tablecomments{Each model is run with five different Gaussian random-phase turbulent velocity field realizations. The quantities given in the rightmost six columns quote the median and full range over these realizations. $\varepsilon_{*, {\rm nfb}}$ indicates the final SFE reached in simulations run with the same initial conditionas over the same time without any stellar winds.}
%\tablenotetext{a}{Only highest resolution is provided.}}
\end{deluxetable*}

\subsection{Simulation Description}
\label{subsec:simulations}

We run a set of simulations of initially uniform density spherical clouds of radius $\Rcld$ and mass $\Mcld$ placed at the center of a cubic domain with side length $\Lbox = 4\Rcld$. Our standard simulations are run on a uniform grid with $\Lbox/\Delta x = 256$, though we explore the effects of resolution in \autoref{app:resolution}. In each simulation the cloud is initially surrounded by gas that is 1000 times less dense and in pressure equilibrium with the cloud, contributing about 1\% of the cloud mass to the total simulation domain. We initialize a  Gaussian random field for the turbulent velocity, with power spectrum $|v_k|^2 \propto k^{-4}$ for wave modes $2\leq k \Lbox/2\pi \leq 64$. The normalization of the velocity field is chosen so that the initial kinetic energy of the cloud is equal to its gravitational binding energy (i.e. a virial parameter of $\alpha_{\rm vir} = 2$).

We set $\Mcld = 10^5 \Msun$ for all of our simulations, and consider models with four different cloud radii, $\Rcld = 20,\, 10,\, 5,\, $ and $2.5\pc$, denoted {\bf R20}, {\bf R10}, {\bf R5}, and {\bf R2.5}. For each cloud radius we run five simulations with different random-phase turbulent initial conditions. Our parameter space spans nearly 3 orders of magnitude in initial density, enabling us to investigate the regulation of star formation by stellar winds for conditions ranging from typical Milky Way GMCs \citep{HeyerDame15,Evans21} to the birth clouds of SSCs \citep{Johnson15,Oey17,Turner17,Leroy18,Emig20}. Each simulation is run (at least) until only 5\% of the cold gas mass remains on the grid, corresponding to a median duration over the turbulent realizations of 14.8, 6.14, 3.49, and 1.75 Myrs for model {\bf R20}-{\bf R2.5}, respectively. 

For the purpose of setting stellar wind feedback through the SB99 code, we adopt an IMF consisting of stepped power laws $dN/dM \propto M^{-\alpha}$.  Our main simulation follow a standard Kroupa IMF with the lowest mass range omitted, that is $\alpha = 1.3$ for $0.1<M/\Msun<0.5$ and $\alpha = 2.3$ for $0.5<M/\Msun<100$ \citep{KroupaIMF}. We have also experimented with a second, top-heavy IMF which has $\alpha = 1.8$ for the higher mass range, but is otherwise identical. Except as noted, all models shown adopt a standard IMF.

\begin{figure*}%
    \centering
    \includegraphics{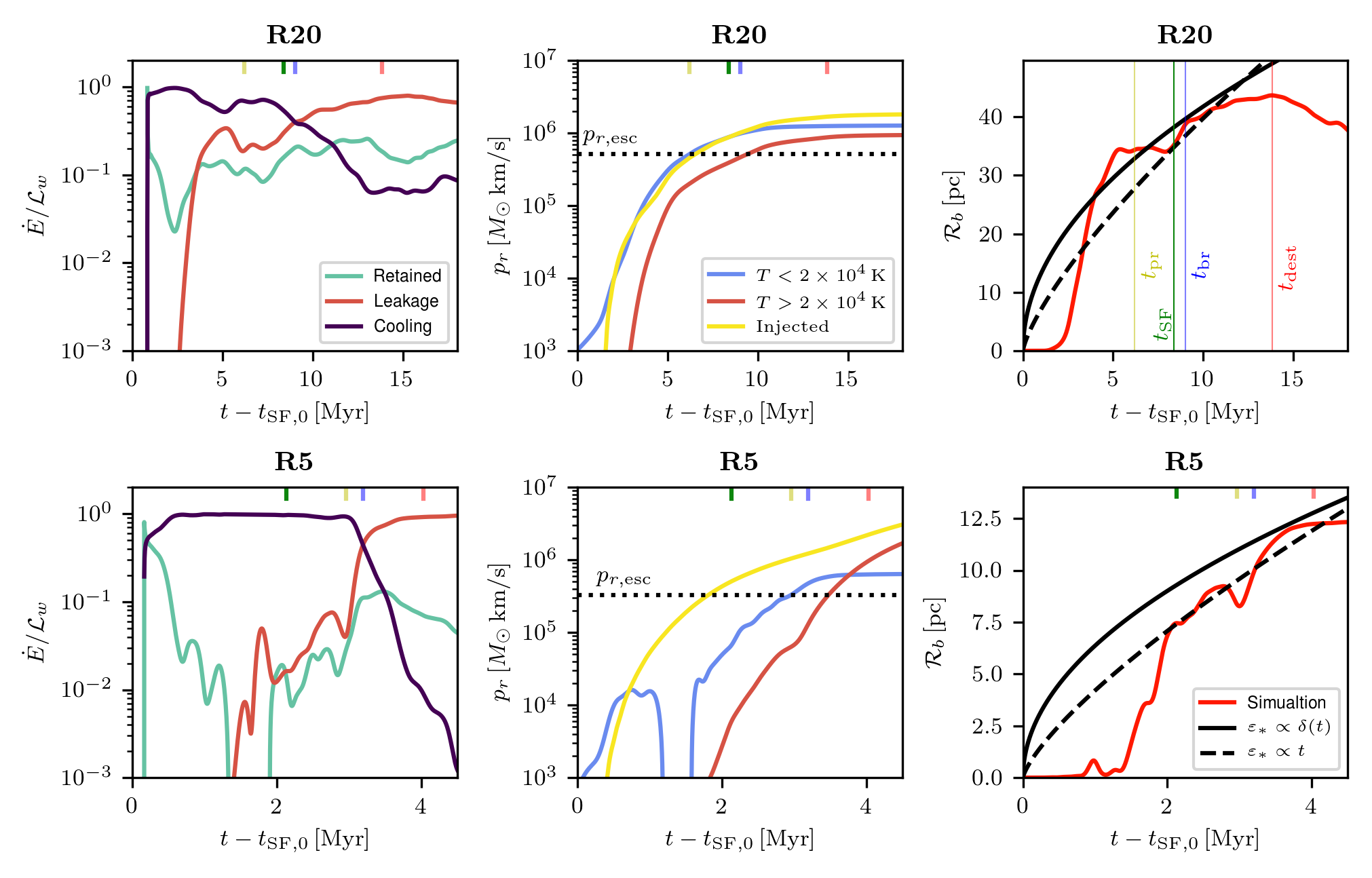}
    \caption{
    Evolution of the energy, momentum, and bubble radius from model {\bf R20} (top row) and {\bf R5} (bottom row) for the same simulations shown in \autoref{fig:snapshot}.The time axis is shifted to coincide with the formation of the first star particle in each model. The ticks at the top of each panel (vertical thin lines in the upper right panel) indicate certain key times: $t_{\rm pr}$, when enough momentum to disperse the cloud has been acquired (yellow, see \autoref{eq:presc_def}); $t_{\rm SF}$, when 95\% of the stellar mass has formed (green); $\tbr$, when breakout occurs (blue); and $t_{\rm dest}$, when the cloud has been effectively destroyed (fuchsia). \textit{Left Column}: The breakdown of input wind energy ``spending'' over time, normalized by the instantaneous wind luminosity. Losses due to cooling (dark purple), bulk advection off the grid (red), and the luminosity fraction retained on the grid (teal) are shown. \textit{Middle Column}: The total radial momentum in wind-polluted ($\fwgrd>10^{-4}$) gas for the cold/warm component (blue), the hot component (red), and the equivalent momentum injected by winds for a single source (yellow). The reference radial momentum required to disperse the cloud ($\presc$) is indicated as a horizontal dotted line. \textit{Right Column}: The effective bubble radius for the simulations, as defined in the text (red lines), with reference predictions for delta-function and linear SFHs (solid and dashed black lines respectively).}
    \label{fig:energy_loss}
\end{figure*}

\section{Results}
\label{sec:results}

In \autoref{fig:snapshot} we show snapshots of the simulations in gas surface density and gas temperature at the time when 75\% of the total stellar mass has formed, corresponding to $t-t_\textrm{SF,0}= 5.5,\, 3.0,\, 1.5,$ and $0.6\, {\rm Myrs}$ for the \textbf{R20}, \textbf{R10}, \textbf{R5} and \textbf{R2.5} models, respectively. Notably, cooling is so efficient in the high surface density clouds that even though a very large stellar mass has accumulated and produced winds, there is still no significant volume of hot gas on the grid.

\subsection{Hot gas losses and cloud evolution}
\label{subsec:lost_energy}

In the high-density regions where star clusters form, the pressure from shocked winds would lead to a momentum input rate exceeding that from radiation feedback unless some of the wind energy is lost \citep{KMBH19,Lancaster21a}.  Two mechanisms for wind energy loss are bulk advection of hot gas out of the dense cloud (``leakage,'' as advocated by \citet{HCM09}), or turbulent interface mixing followed by cooling  \citep[explored in depth in][]{Lancaster21a,Lancaster21b}. Here we show that both effects are important, but at different times. 

In this section we shall consider evolution of energy, momentum, and bubble size, focusing on the {\bf R20} and {\bf R5} models, which are representative of the trends that we see in the other models. 

\subsubsection{Energy Evolution}
\label{subsubsec:energy_pic}

From our simulations, we can measure the total instantaneous mechanical luminosity in winds, $\Lwind$, the net radiative cooling rate in wind-polluted gas\footnote{Similarly to \citet{Lancaster21b} we define wind polluted gas as gas where the fraction of mass in wind material in a given grid cell, $\fwgrd$, is greater than $10^{-4}$.}, $\Ecool$, and the rate at which energy is advected out of the simulation domain in wind-polluted gas, $\Eleak$ (including contributions from thermal and kinetic energy).

Because we are now considering global clouds, we extend the definition of the fractional energy loss rate from \citet{Lancaster21a} to include both cooling, $\Ecool$, and bulk advection out of the domain, $\Eleak$:  
\begin{equation}
    \label{eq:theta_def}
    \Theta \equiv \frac{\Ecool + \Eleak}{\Lwind} \, .
\end{equation}
We refer to $1-\Theta$ as the fraction of wind energy that is retained within the cloud.

In the left column of \autoref{fig:energy_loss} we show the evolution of $\Ecool/\Lwind$, $\Eleak/\Lwind$, and the retained energy fraction ($1-\Theta$) over time in models {\bf R20} (top row) and {\bf R5} (bottom row). Losses transition from being dominated by cooling (at early times) to leakage (at late times). The transition between these two regimes is rapid in the denser models. We define the time at which leakage losses dominate cooling losses as the ``breakout time" $\tbr$.

The low fraction of retained energy indicated by the left column of \autoref{fig:energy_loss} is consistent with the picture, as previously demonstrated in \citet{Lancaster21b}, in which turbulently-mediated energy losses are sufficient to explain the low observed hot-gas pressures in wind-driven bubbles \citep{Rosen14,Lopez11}. The measured hot gas pressures in our simulations are broadly consistent with the predictions of \citet{Lancaster21a}. The deviations (at most $\sim 0.5 \, {\rm dex}$) between the simulations and theory are consistent with moderate momentum enhancement (expected in the theory) or decrement (caused by multiple wind sources, discussed below) along with variations in $\pdotw/\Mst$.

The fraction of energy that remains after cooling and leakage, $1-\Theta$, is only $\sim 1-10\%$, which, during the cooling-dominated period, is consistent with the results of \citet{Lancaster21a,Lancaster21b}. For the denser models, other than the earliest times, $1-\Theta$ is largest after $\tbr$. As we shall show, this is due to an expansion in the volume filled by hot gas.

\subsubsection{Momentum Evolution}
\label{subsubsec:momentum_pic}

The middle column of \autoref{fig:energy_loss} shows the evolution of the gas radial momentum, measured with respect to the center of the simulation domain.  We separate into a cold/warm component ($T<2\times10^4\, {\rm K}$), which is mostly cloud gas, and a hot component ($T>2\times 10^4 \,{\rm K}$), which mostly consists of the expanding wind bubble.  For both components, measured momentum includes contributions from gas on the grid as well as wind-polluted ($\fwgrd > 10^{-4}$) gas that has been advected out of the domain. We also show the total outward momentum that has been injected by winds in the simulations for an equivalent single central source ($\int \pdotw(t) dt$). This total equivalent injected momentum can significantly exceed the actual momentum in the gas, as we discuss further below.

For reference we mark the radial momentum needed to disperse the gas remaining after star formation, $\presc$, which we define as 
\begin{equation}
    \label{eq:presc_def}
    \presc \equiv \Mcld (1 -\sfef)v_{\rm esc} 
\end{equation}
where $\sfef$ is the final SFE of the simulation and $v_{\rm esc}\equiv \sqrt{2G\Mcld/\Rcld}$ is our proxy for the escape speed from the cloud.

\autoref{fig:energy_loss} clearly shows that most of the radial momentum in the cold/warm gas is accumulated before $\tbr$. That is, the time $t_{\rm pr}$ when the radial momentum held in cold/warm gas exceeds the reference ``escape'' value of \autoref{eq:presc_def} is prior to $\tbr$. For the {\bf R5} model, the momentum in the hot gas continues to increase after $\tbr$, at a much higher rate than during the period when momentum was mostly stored in cold/warm gas. This likely does not occur in the {\bf R20} model due to the decrease of $\pdotw/\Mst$ with the age of the stellar population.

In the analytic model of \citet{Lancaster21a} we allowed for the momentum added to the shell to be enhanced above the input wind value $\dot p_w$ by a factor $\alpha_p$. By comparing the total momentum input by winds (yellow lines) to the momentum that is actually accumulated in the gas, we can attempt to assess to what degree momentum is accumulated beyond the ``Efficiently Cooled" solution ($\alpha_p >1$). It is important to note that in our current simulations there are additional effects that can change our interpretation of $\alpha_p$ when performing the comparison we just described. For example, since stars form in over-dense structures, gravity (and the negative momentum associated with collapse) can act to ``crush'' expanding bubbles. This could contribute to accumulated momentum lower than the total input value, appearing as  $\alpha_p < 1$. At the same time, we are not measuring radial momentum with respect to a fixed source, which can be especially confounding at early times when a ``super-bubble'' enveloping the multiples sources has not yet formed. In addition, the contribution from background momentum at early times can be non-negligible. The combination of all of these effects could lead to either $\alpha_p < 1$ or $\alpha_p >1$ at early times depending on the realization.

With these caveats in mind, we can see by comparison of the momenta in \autoref{fig:energy_loss}, that in the {\bf R20} model we quickly reach $\alpha_p \approx 1$. This is consistent with the ``Efficiently Cooled'' solution with some moderate build up of excess momentum when considering the caveats above.

In the dense {\bf R5} model, however, it is apparent that $\alpha_p \lesssim 1$, which is also true for the {\bf R2.5} model. This behaviour can be due both to the effects of gravity described above (which are especially strong in the dense models) and another effect. In the dense clouds, when wind bubbles/shells from individual sources collide with one another at small scales, momentum built up in shells can cancel and the corresponding kinetic energy thermalizes. The swept-up shells in these dense models are dense enough to cool efficiently; with reduced pressure, there is no re-acceleration. This results in a bubble evolution that is technically \textit{less than momentum-conserving}.

\subsubsection{Bubble Radius Evolution}
\label{subsubsec:bubble_radius}

In the right column of \autoref{fig:energy_loss} we show the temporal evolution of the effective radius of the wind-driven bubble, $\reff$, defined as
\begin{equation}
    \label{eq:reff_def}
    \reff = \left(\frac{3V_b}{4\pi} \right)^{1/3} \, .
\end{equation}
Here $V_b$ is the volume of the wind driven bubble, defined as the total volume of cells with temperature $T > 2 \times 10^6 \, {\rm K}$ \textit{or} with radial velocity $v_r > \Vw /2$ as measured with respect to the origin of the simulation domain. Note that in the simulations the bubble is not always a single entity but rather an amalgamation of bubbles from each source which start to expand individually before percolating to become a larger ``super-bubble.'' We use a single effective bubble radius here to give the simplest phenomenological comparison to our previous models, which assume a single, central source. We also sum over all wind inputs in making a prediction for the bubble radius.

For reference, in the right column of \autoref{fig:energy_loss} we show the expected bubble radial evolution for the ``Efficiently Cooled'' (momentum-driven) solution for two different assumed SFHs: 1) An instantaneous, delta-function SFH, which assumes that the final SFE is reached as soon as the first star forms (solid lines) and 2) a linear SFH given by 
\begin{equation}
    \label{eq:lin_sfh}
    \sfe(t)= \frac{\Mst(t)}{\Mcld} = \sfeff 
    \frac{t}{\tffo}
\end{equation}
where $\tffo$ is the initial free-fall time of the cloud and the SFE per free-fall time, $\sfeff = 0.25$, is set based on a fit to the early-time SFHs of the simulations (dashed lines). The first method will tend to over-predict the radial size since at early times it overestimates the stellar power driving the wind bubble. In principle, the latter model should work better, since it more closely follows the actual SFH, and indeed \autoref{fig:energy_loss} shows somewhat better agreement of this model with the numerical results at the time of breakout.  Neither model accounts for the effects of gravity (which are most important in the denser clouds), however, or for the varying strength of the stellar winds with age (which are most important in the longer lived, larger clouds).

\subsection{Star Formation Efficiency}
\label{subsec:predictions}

Like other forms of feedback, winds can help to suppress star formation, but their effectiveness can be reduced by energy losses. Having established that wind energy is efficiently lost through both cooling and leakage, it is interesting to investigate the role winds play in suppressing star formation. To do this, we make use of theories that have been proposed in the literature to predict SFEs subject to feedback mechanisms in the form of momentum injection rate $\dot{p}$. In previous work that accounts for cloud inhomogeneity, this momentum has been conceived of as associated with radiation, but the physical picture of competition between radially outward forces from feedback and inward forces from gravity is quite general, and can be equally well applied to momentum injection from a wind.

Since we evolve our simulations until most of the gas is either driven off the grid or locked in star particles, a straightforward way to assess star formation regulation is to measure the lifetime SFE 
\begin{equation}\label{eq:epsf}
    \sfef \equiv \frac{M_{\rm *,final}}{\Mcld}.
\end{equation}
As a reference value, in \autoref{tab:sim_params} we also give the SFE reached in simulations run for the same time period but without any feedback from stellar winds, which we denote $\varepsilon_{*,{\rm nfb}}$.  Comparison between the simulations with and without feedback shows that at the highest surface density, only a few percent of the mass in the cloud is driven out by feedback.  In contrast, half of the cloud's mass loss can be attributed to feedback in the low-density cloud.

We shall compare the measured SFE in the simulation as a function of cloud parameters to several different theoretical predictions. All of the theoretical models explored here will assume that the wind bubble is in the ``Efficiently Cooled'' limit in which there is no enhancement of momentum relative to the initial wind injection. That is, we assume $\alpha_p = 1$, even though, as we have seen, it can be either slightly greater or less than one depending on the environment.

In our predictions, we will use the mechanical luminosity of the winds per unit stellar mass
\begin{equation}
    \label{eq:Psiw_def}
    \Psi_w \equiv \frac{\Lwind}{\Mst} = %\frac{\pdotw \Vw}{2\Mst} \,  ,
    \frac{\dot M_w V_w^2}{2 \Mst}
\end{equation}
and the total radial momentum input rate for an equivalent single central source,
\begin{equation}
  \pdotw = \frac{2 \Psi_w}{V_w} \Mst;   
\end{equation}
here $\Lwind$ is the total mechanical luminosity of all winds collectively, and $\Vw$ is the wind injection velocity. For our analytic predictions below, we simplify by using the initial value of $\Psi_w$ determined by the SB99 code for a fully-sampled Kroupa IMF, which is approximately $10^{34} \, {\rm erg/s/}M_{\odot}$. We note that in detail, the IMF-averaged $\Psi_w$  stays relatively constant until an increase due to Wolf-Rayet Stars around $t \approx 3 \, {\rm Myr}$ and subsequent decline beginning at $t \approx 4 \, {\rm Myr}$ (see Figure 3 of \citet{Lancaster21a}). While our numerical models employ a time-dependent IMF-averaged value of $\Psi_w$ for each source, our analytic models adopt this simpler approach of constant $\Psi_w$.

Suppression of star formation in clouds occurs by driving mass out of the system. We consider a fluid element  of mass $M$ and cross-sectional area $A$ at a distance $r$ from a central cluster. If it  absorbs all incident momentum $ \dot p = \pdotw A/(4\pi r^2)$ from a radially-flowing wind, the Eddington ratio between the wind force $\dot p$ and the gravitational force $G M_* M/r^2$ of the cluster on the fluid element is 
\begin{subequations}
\begin{eqnarray}
\label{eq:Gammaw_def}
    \frac{\dot{p}}{G M_* M/r^2} &=& \frac{\Psi_w}{2\pi G \Vw\Sigma } \equiv \frac{\SEdd}{\Sigma} \\
    &=& 0.57 \left(\frac{\Psi_w}{10^{34} \, {\rm erg/s/}\Msun} \right)
    \left(\frac{\Sigma}{10^3 \, \Msun\, {\rm pc}^{-2}} \right)^{-1} \nonumber\\
    & &\times \left(\frac{\Vw}{10^3 \, {\rm km/s}} \right)^{-1}\, .
\label{eq:Gammaw_defb}
\end{eqnarray} 
\end{subequations}
Here, the gas surface density of the fluid element is $\Sigma = M/A$, and we have defined the  Eddington surface density $\SEdd \equiv \Psi_w/(2 \pi G V_w)$ as the maximum value for which the  feedback wind momentum exceeds gravity; for our adopted parameters this is $\SEdd=570 M_\odot\ \pc^{-2}$.

\subsubsection{Lancaster et al. (2021a) Prediction}
\label{sec:lancastersfe}

We begin by reviewing the simple prediction of  \citet{Lancaster21a} for the SFE, obtained under the assumption that all star formation occurs instantaneously at the center of a cloud, creating a cluster of mass $\Mst = \sfe \Mcld$, and that the rest of the mass $(1-\sfe)\Mcld$ is swept into a spherical shell at the cloud radius $\Rcld$. For the force from the wind to exceed the gravity of the cluster plus the self-gravity of the shell, the condition
\begin{equation}
    \pdotw > \frac{G\Mcld^2}{\Rcld^2} \left(\sfe(1-\sfe) + \frac{(1-\sfe)^2}{2} \right)
\end{equation}
must be satisfied. The above can be written as
\begin{equation}
    \label{eq:lancaster21a_condition}
    8\frac{\SEdd}{\Scld} > \frac{1-\sfe^2}{\sfe} \, ,
\end{equation}
where $\Scld \equiv \Mcld/\pi\Rcld^2$. This inequality may be solved to obtain the minimum $\sfe$ given $\SEdd/\Scld$.  For low  $\Scld $, i.e. $\Scld \ll \SEdd$, this yields $\sfe\approx \Scld/(8 \SEdd) \propto \Scld$ , while for high $\Scld $, $\sfe \approx 1- 4 \SEdd/\Scld$  \citep{Lancaster21a}.

The above parallels simple scaling arguments for SFE as limited by radiation pressure, where instead of $\pdotw$ the radiation momentum injection $L/c$ (for $L$ the bolometric luminosity) is used \citep[e.g.][]{Fall10,Murray11,Raskutti16,Li_Vogelsberger2019}. The assumption of instantaneous star formation is, however, clearly inconsistent with the more realistic simulations carried out here.  

\subsubsection{Steady Star Formation}
\label{sec:steadysfe}

An alternative approach allows for the SFH to be extended in time, rather that instantaneous, while still occurring at the center of the cloud. Following the same derivation for bubble expansion as in \citet{Lancaster21a} (ignoring the effects of gravity and assuming a constant momentum input rate per unit stellar mass $\pdotw/\Mst$), while setting $M_* /M_{\rm cl} = \sfe  = \sfeff t/\tffo$, the bubble expands as 
\begin{equation}
    \frac{\reff}{\Rcld} = \left[\frac{16\SEdd}{3\Scld\pi^2} \sfeff \left(\frac{t}{\tffo}\right)^3 \right]^{1/4} \, ,
\end{equation}
where we have defined the initial free-fall time of the cloud as 
\begin{equation}
    \label{eq:tff_def}
    \tffo \equiv \sqrt{\frac{3\pi}{32 G \rhobar}} = 
    \sqrt{\frac{\pi^2 \Rcld^3}{8G\Mcld}}\, .
\end{equation}

If we then assume that star formation is immediately halted once the bubble reaches the edge of the cloud, we can obtain an estimate for $\sfe$ by setting $\reff = \Rcld$ in the above and solving for $t$.  This gives us the final SFE $\sfef =[3 \pi^2 \sfeff^2 \Scld/(16 \SEdd)]^{1/3}$ for an assumed $\sfeff$.

This model may be accurate in large, low-density clouds where the effects of gravity on the wind bubble expansion are actually quite small. However, there is no consideration for the ability of the swept-up shell to escape the gravity of the system, and if the bubble expands slowly enough this model will predict $\sfe>1$, which is clearly unphysical.

\subsubsection{Thompson \& Krumholz (2016) Prediction}
\label{sec:thompsonsfe}

The models of \autoref{sec:lancastersfe} and \autoref{sec:steadysfe} assume the driving source ``sees'' only a single gas surface density over all solid angles, given by the mean value for the cloud as a whole. In reality, since the clouds in which stars are born are supersonically turbulent, the surface densities viewed either externally or looking outward from the center will have a log-normal distribution \citep[e.g.][]{Raskutti17,JGK19}. Considering momentum injection associated with radiation pressure, \citet{ThompsonKrumholz16} pointed out that directions where the total line-of-sight surface density is sufficiently low can be ``super-Eddington'' even if the cloud as a whole is sub-Eddington. Along these lines of sight, gas may be driven out of the cloud.  

\citet{ThompsonKrumholz16} denote the  mass-weighted distribution of (logarithmic) surface density $x\equiv \ln \left(\Sigma/\Savg\right)$ by $p_-(x)$, where $\Savg \equiv \int \Sigma d \Omega/4\pi$ is the sky-averaged surface density and the mean of $p_-(x)$ is $\slnS^2/2$.
Integrating over the distribution, 
\begin{subequations}
\begin{eqnarray}
\label{eq:zeta_def}
    \zeta_{-}(x) &\equiv& \int_{-\infty}^x p_{-} (x') dx' \\ 
    &=& \frac{1}{2} \left[ 1 -{\rm erf} 
    \left(\frac{-2x+ \slnS^2}{2\sqrt{2}\slnS} \right) \right] \, .
\label{eq:zeta_defb}
\end{eqnarray} 
\end{subequations}

Given a distribution of mass, \citet{ThompsonKrumholz16} posit an instantaneous loss rate of mass ejected from the cloud as 
\begin{equation}
    \label{eq:TK_windMdot}
    \dMclw = \zeta_-(x_{\rm crit}(t)) 
    \frac{M_g (t)}{\tff (t)} \, ,
\end{equation}
where $M_g(t)$ and $\tff(t)$ are the instantaneous gas mass and free-fall time of the system\footnote{$\tff(t)$ is defined here as in \autoref{eq:tff_def} except with $\Mcld$ replaced by the sum of the instantaneous gas, $M_g(t)$, and stellar, $\Mst (t)$, mass.}, and $x_{\rm crit}$ is determined by the condition for a structure to be super-Eddington. At the same time, star formation is assumed to follow
\begin{equation}
    \label{eq:lin_sfh_tk}
    \dot{M}_* = \sfeff \frac{\Mgas(t)}{\tff (t)} \, ,
\end{equation}
where $\sfeff$ is a constant SFE per free-fall time, so that allowing for mass loss to both outflow and star formation (i.e. $\Mgas(t) = \Mgas(0) - \Mst(t) - \Mclw(t)$), the average surface density evolves in time as $\Savg = 4\Rcld(t) \rhobar(t)/3$.  \citet{ThompsonKrumholz16} allow the cloud radius to evolve as well, but we keep this fixed.  

Adapting from the case of radiation feedback to wind feedback, the limit in the integral over the distribution becomes $x_{\rm crit}= \ln (\Sigma_{\rm TK}/\bar\Sigma)$ where $\Sigma_{{\rm TK}} =\SEdd/(1 + M_g/\Mst)$ is the Eddington surface density as defined in \citet{ThompsonKrumholz16}, allowing for gravity of both stars and gas.  

The above equations can be integrated jointly forward in time until all mass that was originally in gas is either in wind or stars. The final stellar mass then determines the final predicted SFE.  This is a function of the initial cloud properties, the adopted $\Psi_w$, and the values assumed for the parameters $\sfeff$ and $\slnS$. 

\subsubsection{Raskutti et al. (2016) Prediction}
\label{sec:raskuttisfe}

\citet{Raskutti16} also developed a prediction for the SFE of a cloud regulated by radiation pressure feedback that takes in to account the log-normal distribution in surface densities seen by a source particle. The theoretical model is very similar to that of \citet{ThompsonKrumholz16} except that, instead of integrating ODEs they consider the instantaneous state of a cloud.  If some fraction, $\sfe$, of the original cloud mass has been gathered into a central star cluster and the remaining gas is distributed around this source with a log-normal distribution in surface densities, they calculate the fraction of the original cloud that would be super-Eddington as
\begin{equation}
    \eof(\sfe) = (1 - \sfe) \zeta_-(x_{\rm crit}(\sfe)) \, ,
\end{equation}
where the only differences between this $x_{\rm crit}$ and that of \citet{ThompsonKrumholz16} are (1) a factor of $1/2$ in the gas gravity term, and (2) the inclusion of a parameter to account for the possible contraction of the cloud as it evolves (which exists to a different degree in the \citet{ThompsonKrumholz16} model).

\citet{Raskutti16} argue that star formation proceeds until a SFE of
\begin{equation}
    \label{eq:raskutti_min}
    \varepsilon_{{\rm min }} = \argmax_{0<\sfe<1} \eof(\sfe)
\end{equation}
is reached. At this point, the maximum fraction of the original gas satisfies the Eddington condition and will (in their model) be immediately ejected. This corresponds to a minimum prediction for the SFE, since there is still some gas left in the shell at this point that is sub-Eddington. If all of this remaining gas collapses to form more stars then the final SFE will instead be
\begin{equation}
    \label{eq:raskutti_max}
    \varepsilon_{{\rm max}} = 1 - \eof(\varepsilon_{{\rm min}}) \, .
\end{equation}
To obtain SFE predictions (upper and lower bounds) a value for the parameter $\slnS$ must be adopted, but no assumption for $\sfeff$ is needed.

\subsubsection{Comparison to Simulations}
\begin{figure}
    \centering
    \includegraphics{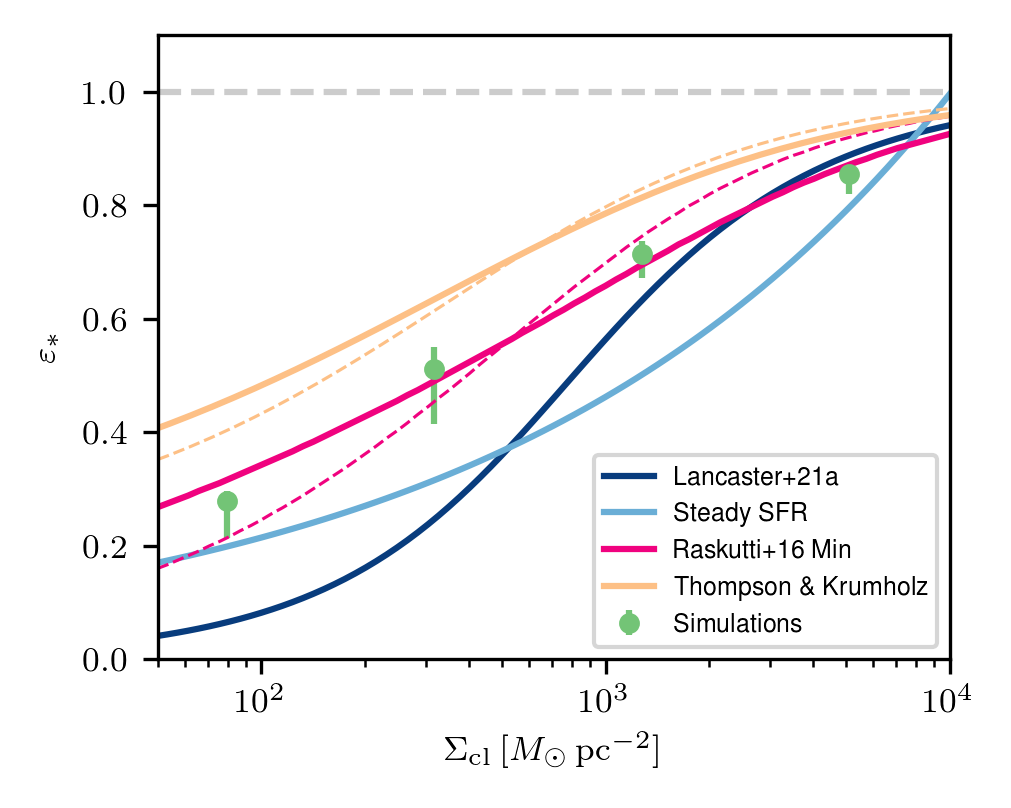}
    \caption{The final SFE realized in our simulations in comparison to analytic models.
    %with a standard IMF, at each simulated cloud radius 
    For each simulation parameter set (see \autoref{tab:sim_params}) we show the median SFE as a point, with the range of the SFEs from all runs shown by the vertical error bars. Predictions for the SFE based on several different models are shown, as labeled. For lognormal-based models, solid lines denote $\slnS = 1.5$ and dashed lines denote  $\slnS = 0.75$ (see text).}
    \label{fig:sfe_scaling}
\end{figure}

In \autoref{fig:sfe_scaling} we show the final SFEs for all of our simulations, displaying results based on the five different runs for each parameter set. For comparison we show the predictions of each of the analytic models described in the previous sections.

The assumptions and parameters used for each of these theories are as follows
\begin{itemize}
    \item[$\bullet$] \textit{Lancaster+21a}: The only free parameter is the momentum input rate per unit mass $\pdotw/\Mst= 2\Psi_w/\Vw$. We adopt a value $8.6 \, {\rm km/s/Myr}$, determined using the average over the first 1 Myr using  SB99 \citep{SB99}. We use this value of $\pdotw/\Mst$ in all models.
    
    \item[$\bullet$] \textit{Steady SFR}: We adopt $\sfeff = 0.25$, broadly consistent with the SFH in our simulations. 

    \item[$\bullet$] \textit{Thompson \& Krumholz}: As above we adopt  $\sfeff = 0.25$, and for the width of log-normal in surface density we adopt $\slnS = 1.5$ as a value consistent with many of the simulations over much of the evolution. 

    \item[$\bullet$] \textit{Raskutti+16}: We display the minimum prediction given by \autoref{eq:raskutti_min}, also assuming  no expansion or contraction of the cloud ($x=1$ in their notation). We adopt $\slnS = 1.5$ as above.
\end{itemize}

We first note that the agreement between the \citet{Lancaster21a} prediction and our simulations is best for the highest surface density clouds. This is because the SFH most closely resembles an instantaneous starburst in the high-$\Sigma$ regime, for which the star formation timescale is much less than the cloud disruption timescale (as we will see in \autoref{subsec:timescales}). However, the \citet{Lancaster21a} prediction significantly underestimates the measured SFE at low surface densities. 

The steady star formation prediction does reasonably well at low surface densities, but quite poorly elsewhere. This is because the condition chosen to turn off star formation is arbitrary (when the bubble reaches the edge of the cloud) and does not take into account the effect of gravity in any way. There is also no built-in limit of star formation to prevent $\sfe$ from exceeding unity, as it does for $\Scld \gtrsim 10^4 \, \Msun {\rm pc}^{-2}$.

The \citet{ThompsonKrumholz16} model does a much better job at capturing the overall trend in the simulation results, while still slightly overestimating the measured SFE. In principle the model could be made to fit better by adjusting its parameters, but our specific choices for the parameters are motivated by the simulations themselves.

The \citet{Raskutti16} model, which has no explicit time dependence and assumes a constant PDF in $\Sigma$, does reasonably well at high $\Scld$ and slightly over-predicts $\sfe$ at low $\Scld$. The latter could be due to a break-down of the single, centralized source or constant $\pdotw/\Mst$ assumptions (see below).

Since in practice $\slnS$ is not constant in time, we additionally show the predictions of both the \citet{ThompsonKrumholz16} and \citet{Raskutti16} models for $\slnS=0.75$ in \autoref{fig:sfe_scaling} as thin dashed lines. This value is chosen as comparable to the measured PDF variance at the onset of star formation.  Adoption of this value leads to a slightly higher (lower) predicted SFE at higher (lower) $\Sigma$ for both models. However, we caution that there is not a strong physical motivation for preferring this choice.  Rather, the comparison provides a feeling for the parameter sensitivity of the two models.  

As mentioned above, we have also investigated the effects of top-heavy IMFs in our denser models ({\bf R5} and {\bf R2.5}), motivated by observations of star clusters in dense environments \citep{McCrady05,Lu13,Hosek19}.  We find that the greater strength of the winds (larger $\pdotw/\Mst$) with a top-heavy IMF is noticeable, reducing the final SFEs from 71 to 53\% and 85 to 75\% in the {\bf R5} and {\bf R2.5} models respectively.

Finally, as we alluded to above, we note that the approximation of $\pdotw \approx {\rm const.}$ breaks down once the Wolf-Rayet stage of stellar evolution occurs around $\sim 3\, {\rm Myr}$ (see \citet{Lancaster21a} Figure 3), with $\pdotw$ increasing by a factor $\sim 3$ and then declining again over the next few Myr. As we shall show in the next section, the two lower-density models have lifetimes that extend into the Wolf-Rayet phase, so one might expect slightly lower $\sfe$ than predicted under the model assumption of constant $\pdotw/\Mst$, as is indeed the case. Because their lifetimes are shorter, the higher $\Scld$ simulations would not be significantly affected by the increase of $\pdotw$ in the Wolf-Rayet stage.

\begin{figure}[t!]
    \centering
    \includegraphics{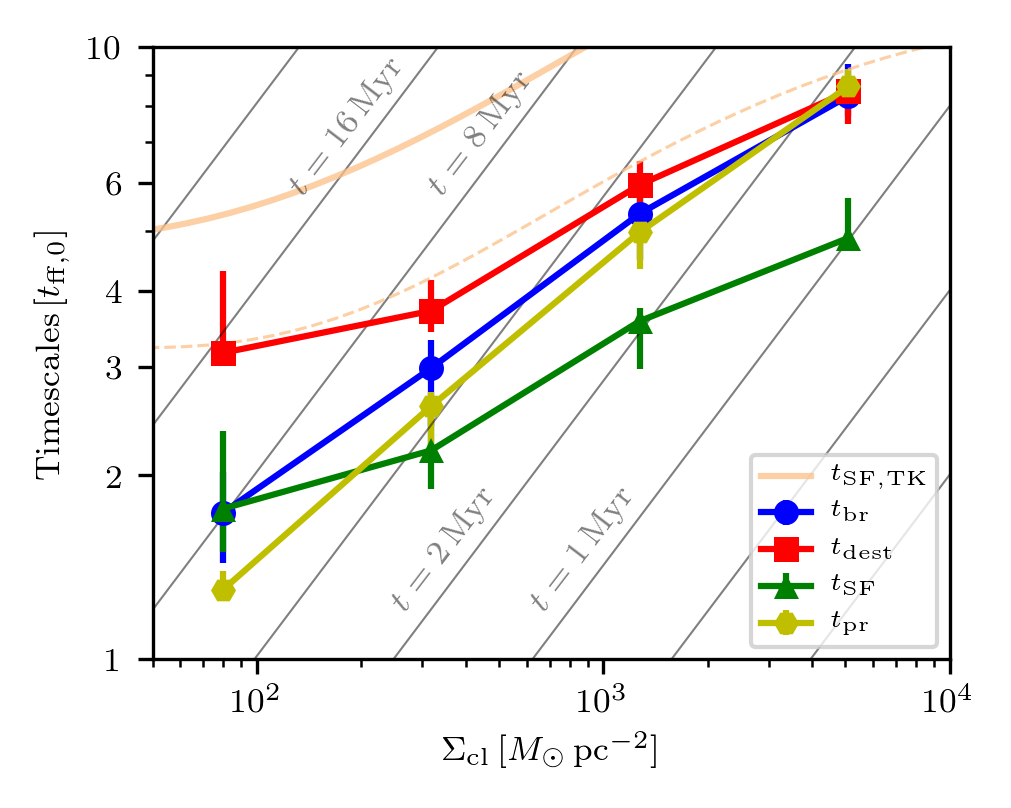}
    \caption{Timescales of key events, as measured in the simulations (see \autoref{subsec:timescales} for definitions);  values are shown in units of the free-fall time. Lines of constant physical time are also shown as thin grey lines. The prediction of  \citet{ThompsonKrumholz16} for $t_{\rm SF}$ is shown as beige lines for $\slnS = 1.5$ (0.75) in solid (dashed) line styles.}
    \label{fig:timescales}
\end{figure}

\subsection{Timescales}
\label{subsec:timescales}

In \autoref{fig:timescales} we show various time-scales measured from the simulations, and how they depend on the surface density of the cloud. All timescales represent the time elapsed between the onset of star formation and the physical event described below, and in \autoref{fig:timescales} are plotted in units of the initial free-fall time in the cloud.
\begin{itemize}
    \item[ $t_{\rm br}$:] The ``breakout time'' is when energy losses from the wind become dominated by bulk advection (leakage) of hot gas out of the cloud, rather than turbulent mixing and subsequent cooling (see \autoref{subsubsec:energy_pic}).

    \item[$t_{\rm SF}$:] The ``star formation time'' is when 95\% of the final stellar mass has been formed in the simulation.

    \item[$t_{\rm dest}$:] The ``destruction time'' is when  the gas mass remaining on the grid drops to 5\% of the initial cloud mass.
    
    \item[$t_{\rm pr}$:] The ``radial momentum time'' is when the total radial momentum carried in the cold/warm ($T<2\times 10^4 \, {\rm K}$) gas exceeds the ``escape momentum'' defined in \autoref{eq:presc_def}.
\end{itemize}

Considering the various timescales together, \autoref{fig:timescales} reveals a common sequence in which winds halt star-formation while cooling is still dominating the wind's energy losses and acceleration is still occurring, but the winds eventually break out from the dense cloud gas, and the remaining gas is gradually driven out of the simulation domain. The momentum needed to disrupt the cloud is always accumulated in the gas \textit{before} advection begins to dominate over cooling losses ($t_{\rm pr} < t_{\rm br}$).  Interestingly, it is also broadly true that $t_{\rm dest}\approx 2t_{\rm SF}$.

The details of evolution nevertheless vary over the range in surface densities probed by our simulations. At the lowest surface densities probed here, the momentum  needed to eventually disrupt the cloud is acquired \textit{before} star formation halts. In this case, star formation is a continuous process that proceeds in parts of molecular clouds while it is simultaneously being halted in other parts. However, at higher surface density, star formation essentially is halted and then winds engage in an extended battle to turn around and expel lower-density gas while also trying to break out of the cloud. In these denser clouds, $t_{\rm br}$, $t_{\rm pr}$, and $t_{\rm dest}$ all occur roughly simultaneously. For the highest surface density cloud, as we have shown, very little material is in practice driven away by feedback.

\begin{figure*}
    \centering
    \includegraphics{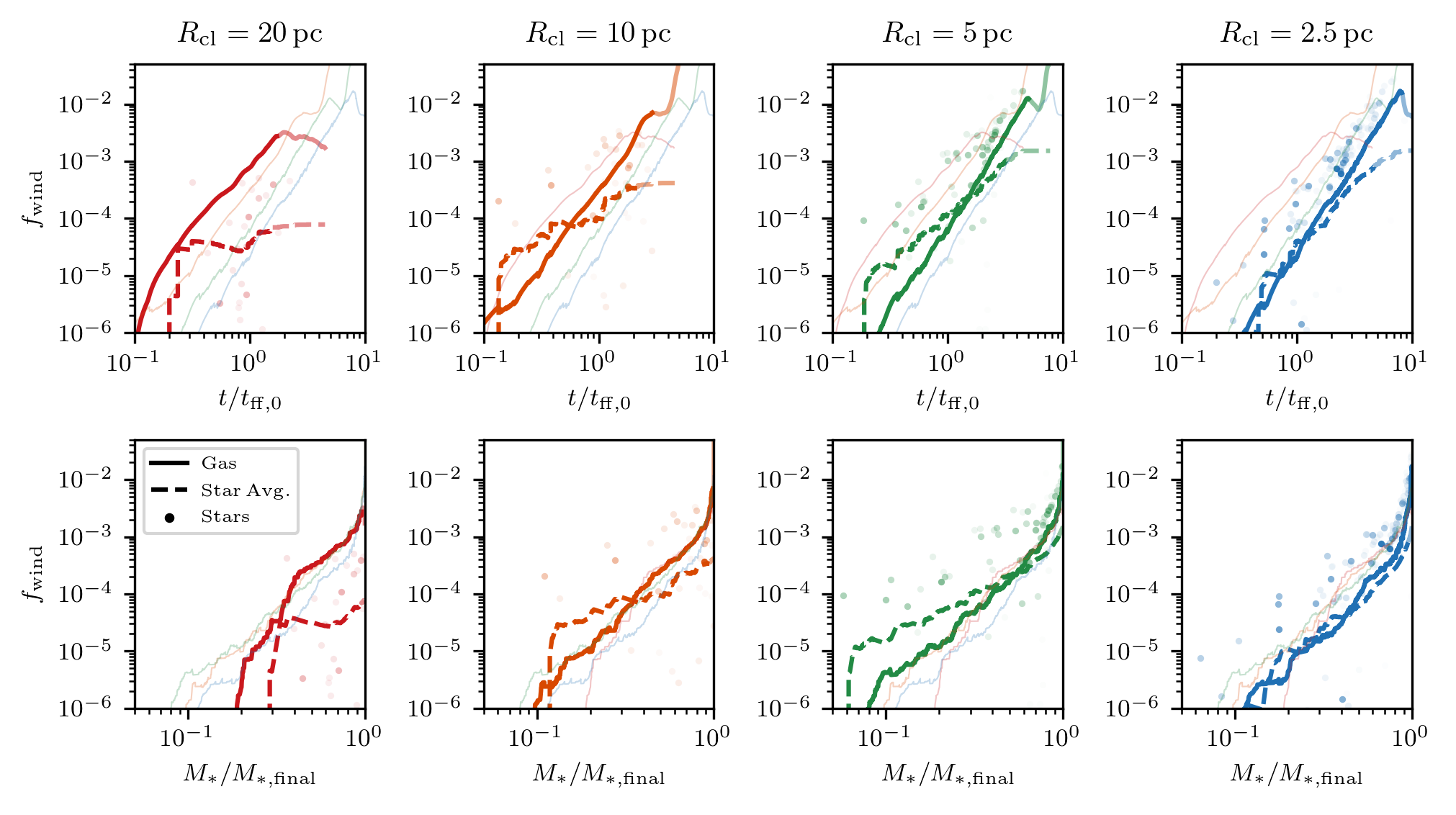}
    \caption{Evolution of the global fraction of wind mass $\fwind$ in the gas (solid lines), in the stars overall (dashed lines), and in individual star particles (points, where more opaque points indicate more massive star particles). We show results as a function of time in units of the initial free-fall time (top row), and as a function of the fraction of the final stellar mass that has formed (bottom row). For the individual star particles, we show the fraction of their \textit{initial} mass in wind material as a function of the time when they first formed (or fraction of stellar mass formed when they first formed, in the bottom row). Each panel also includes $\fwind$ in gas from other models as thin, transparent lines. We show  $\fwind$ for the gas (stars) as opaque curves up until $t_{\rm br}$ ($t_{\rm SF}$), after which more transparent lines are shown. We note that the dashed lines include \textit{all} star particles, even those with negligible enrichment that do not appear on this plot.}
    \label{fig:wind_pollution}
\end{figure*}

Since the model of \citet{ThompsonKrumholz16} provides for a full SFH, it gives a prediction for $t_{\rm SF}$. We include this prediction in \autoref{fig:timescales}. The prediction overestimates the star-formation timescale by a factor of $\approx 2$ for the $\slnS = 0.75$ case (dashed line) and by a larger factor for the $\slnS = 1.5$ value (solid line). The fact that star formation halts much more rapidly than the model predicts indicates that the model assumption of a wind acceleration timescale $\tff(t)$ is not borne out in our simulations.  Potentially, other assumptions for the acceleration timescale (e.g.  accounting for the distribution of Eddington ratios) could be explored, but a reformulation of this theory is beyond the scope of the present work.

\subsection{Wind Pollution}
\label{subsec:wind_pollution}

One of the central goals of this work is to provide a first investigation of the incorporation of wind material in new stars, and its dependence on cloud environment. As a caveat, we note that the winds investigated here, from massive OB stars, are not very chemically distinct from their parent gas, especially before the onset of the Wolf-Rayet phase \citep{Leitherer92}, and thus the ``pollution'' by these winds would not produce compositional signatures. The winds that are thought to self-enrich forming globular clusters presumably would have somewhat different physical properties (e.g. mass-loss rates,  wind velocities), and this difference should be kept in mind when interpreting the results we present.

As discussed in \autoref{sec:methods}, we use a passive scalar, injected with the wind material, to track how wind material is mixed into the surrounding gas and subsequently-forming stars. In \autoref{fig:wind_pollution}, we present the evolution of the \textit{global} fraction of mass, in stars ($\fwnds$) and in gas ($\fwind$), that originated from stellar winds. All simulations depicted in \autoref{fig:wind_pollution} have identical random-phase initialization of their turbulence. 

From \autoref{fig:wind_pollution}, the evolution of $\fwind$ in the gas (solid lines) is remarkably similar across the simulations when the evolution is shown in units of the free-fall time. This is consistent with each of the simulations having nearly the same (linear) SFH at early times with $\sfeff \approx 0.25$, which then implies $\Mw \propto t^2$ (in practice, the slope in the simulations is closer to $\Mw \propto t^{5/2}$). These lines significantly deviate when the wind bubbles break out, first causing $\fwind$ to decrease as much wind material is advected off of the grid and then increase as all remaining, non-wind material is dispersed.

For the denser clouds, the fraction of wind material in star particles (dashed lines) very roughly tracks the evolution of $\fwind$ in the gas, until star formation slows (dashed lines becoming transparent indicate 95\% of star formation is complete). As we have seen in \autoref{fig:timescales}, in denser clouds, star formation takes place over more free-fall times, resulting here in a more self-polluted stellar population.

In the denser clouds, polluted stars tend to form in gas that is preferentially more polluted than the average gas (points fall above the solid lines), indicating that mixing is less efficient in these clouds. This is somewhat counter-intuitive since one might expect that mixing would be more efficient in the denser clouds where the turbulent Mach numbers are higher, promoting a larger degree of turbulence compared to radial expansion of wind bubbles (see Figure 10 of \citet{Lancaster21b}). However, as we have seen, when gravity is considered wind bubbles are much more easily contained in the dense clouds. Indeed when we look at the mass-weighted distribution of $\fwind$ in the simulations, the denser clouds are much less well-mixed, explaining the trend we see here.

To the best of our knowledge, this sort of self-consistent analysis of pollution by the winds from newly-formed stars is the first of its kind that has been shown in the star cluster formation literature. We anticipate that future analyses of this kind, including chemical enrichment from other wind sources, can be used to better understand proposed formation mechanisms for multiple populations in Globular Clusters \citep{BastianLardo18,Gratton19,Bailin18}, as well as chemical evolution within normal, redshift-zero star-forming clouds. However, it is important to keep in mind that there are many potentially important physical phenomena which we have not included that may or may not be important in different scenarios (radiative feedback, magnetic fields, dust, varying metallicity).

\section{Summary and Discussion}
\label{sec:conclusions}

The simulations and analyses presented here explore the applicability of the ``Efficiently Cooled'' stellar wind bubble model \citep{Lancaster21a,Lancaster21b} in the more realistic scenario where stars form self-consistently through gravitational collapse, which competes with cloud dispersal by stellar winds. We show the core principle of the ``Efficiently Cooled'' model holds: enough energy is removed through radiative cooling to make the wind bubbles momentum-driven rather than energy-driven (see \autoref{fig:energy_loss}).

In the absence of losses, stellar winds would dominate other ``early feedback'' in star-forming clouds. The weak observational signatures of winds has led to suggestions that energy is efficiently removed by either (i) bulk advection of hot gas out of the cloud, or (ii) radiative cooling mediated by turbulent mixing. We show that over the time-scales relevant to star formation (see \autoref{fig:timescales}), it is the latter process that is most important for removing energy injected by stellar winds. This conceivably depends upon the exact geometry of the cloud, and it is plausible that for highly porous clouds, bulk advection could dominate energy losses.

We apply theoretical models to make predictions for SFEs in clouds on the basis of ``Eddington ratio-like'' considerations of competition between gravity and wind forces, taking into account turbulence-driven substructure. We compare predictions of these theoretical models for SFEs to the numerical results from our simulations (see \autoref{fig:sfe_scaling}). While our simulations include realistically time-varying specific wind momentum input rate, the theoretical models assume $\pdotw/\Mst$ is constant, for simplicity. Despite this, we find that the models derived from \citet{Raskutti16} follow the numerical results fairly well, while the \citet{ThompsonKrumholz16} models slightly over-estimate the final SFEs. Model assumptions regarding (i) constant momentum input rates, (ii) a constant SFE per free-fall time, $\sfeff$, and (iii) a log-normal distribution of surface densities with constant variance are not fully satisfied, however, which may account for some of the deviations between numerical results and models. We also compare the \citet{ThompsonKrumholz16} prediction of the star formation duration $t_{\rm SF}$ to our numerical results, and find that the former overestimates our measurements of $t_{\rm SF}$, which we link to their assumption for the timescale for gas expulsion.

We do not attempt to make a direct comparison between our simulated SFE and observed values for different cloud regimes, given that important feedback effects are left out in our numerical models. Nevertheless, it is conceivable that stellar wind feedback is the dominant star formation regulation mechanism in very dense clouds \citep{Levy20,Lancaster21a}, and it is interesting to compare our models to observations in this regime. The SFEs we measure at high surface density are broadly consistent with the range inferred from observations of SSC-forming clouds at the centers of galaxies, which are generally $\sfe \gtrsim 70 \%$ \citep{Leroy18,Emig20}. Due to the range of uncertainties associated with the observations, empirical SFEs are also broadly consistent with the SFEs in our simulations that assume stellar populations with top-heavy IMFs, as might be expected in these clouds \citep{Hosek19}.

In typical Milky Way (or LMC/SMC) GMCs, similar to our lower density models, momentum-driven stellar wind feedback is not expected to be the dominant star formation regulation mechanism \citep{Lancaster21a}. However, it is still interesting to quantitatively assess the effects of winds in this regime. Based on the work of \citet{Lopez11,Rosen14} and others, there are large discrepancies between observations of the hot gas pressure produced by winds and the classical \citet{Weaver77} model, and in \citet{Lancaster21a,Lancaster21b} we proposed that energy losses due to turbulently-mediated radiative cooling could explain these discrepancies. Inspection of the hot gas pressure in the more-realistic simulations presented here bear out this proposal, where our {\bf R20} simulations show hot gas pressures ranging from $10^5-10^7\, {\rm K}\,{\rm cm}^{-3}$ at $t\lesssim 3 \,{\rm Myr} $, broadly consistent with X-ray observations of nearby massive star-forming regions \citep{Lopez11,Lopez14}.

Finally, using a passive scalar that is injected with the wind material, we investigate the process of wind pollution. The fraction of the gas originating in winds scales roughly as $\fwind \propto t^{5/2} \propto \Mst^{5/2}$ , and for the densest cloud (model {\bf R2.5}) the mean wind-mass fraction of the stellar population generally follows this up to the time of wind breakout.  For the less-dense clouds the mean pollution of the stellar population increases less steeply in time, and individual star particles can have significantly higher or lower $\fwind$ than the mean value at the time they form.

The increase of $\fwind$ in time, combined with the fact that star-formation takes place over more dynamical times in denser clouds, implies stellar populations with larger fractions of wind material in denser clouds. This investigation opens the door for future direct numerical simulations to explore wind pollution and chemical self-enrichment in cluster-forming clouds in more depth (incorporating additional physical processes and winds that are more chemically enriched), with implications for the study of galactic archaeology and multiple populations in globular clusters \citep{FBH02,BastianLardo18}.

\begin{acknowledgements}

L.L. thanks Erin Kado-Fong, Todd Thompson, and Madeline Lucey for useful comments and references. We would like to thank the referees for useful comments that helped to improve the paper. This work was partly supported by the National Science Foundation (AARG award AST-1713949) and NASA (ATP grant No. NNX17AG26G). J.-G.K. acknowledges support from the Lyman Spitzer, Jr.,~Postdoctoral Fellowship at Princeton University. Computational resources were provided by the Princeton Institute for Computational Science and Engineering (PICSciE) and the Office of Information Technology’s High Performance Computing Center at Princeton University.

\end{acknowledgements}

\software{
{\tt Athena} \citep{Stone08_Athena,StoneGardiner09},
{\tt astropy} \citep{astropy13,astropy18}, 
{\tt scipy} \citep{scipy},
{\tt numpy} \citep{harrisNumpy2020}, 
{\tt IPython} \citep{Perez07}, 
{\tt matplotlib} \citep{matplotlib_hunter07},
{\tt xarray} \citep{hoyer17}, 
{\tt pandas} \citep{pandas2020},
{\tt adstex} (\url{https://github.com/yymao/adstex})
}


%\vspace{5mm}
%\facilities{HST(STIS), Swift(XRT and UVOT), AAVSO, CTIO:1.3m,CTIO:1.5m,CXO}


%\software{astropy \citep{2013A&A...558A..33A}, Cloudy \citep{2013RMxAA..49..137F},SExtractor \citep{1996A&AS..117..393B}}


%\newpage
\appendix

\section{Effects of $\Mcut$}
\label{app:mcut}

In this appendix we explore the effects of varying the prescription for the assignment of stellar wind power, $\Lwind$, to a star particle. All simulations presented in this section have identical conditions to those simulations outlined in columns 2-5 of \autoref{tab:sim_params} but are run at half the resolution ($2\Delta x$ as given in column 6 of \autoref{tab:sim_params}).

It has been shown in the literature \citep[e.g.][]{Grudic19,Smith21} that prescriptions for IMF sampling in imposing feedback can have significant consequences for the key results of a given simulation, despite being a rather arbitrary prescription for physics that is sub-grid to the simulation. To provide some sense of the systematic uncertainty in our results  that arises from ``model variance'' in our simulations, we present here a few numerical experiments covering a small range of different prescriptions for the assignment of feedback to star particles.

The first prescription we use is the same outlined in the text, with $\Mcut = 10^3 \Msun$. We also run simulations here with the same prescription except with $\Mcut = 10^2 \Msun$. Finally, we include a third prescription based on the Poisson sampling method outlined in \citet{Su18} and also used in \citet{Grudic19}. This method works by sampling a number of ``O-stars'' from a Poisson distribution whose mean depends on the mass of the sink particle to which we are assigning feedback. Specifically, for each star particle of mass $\Mstsnk$ we sample from a Poisson distribution of mean
\begin{equation}
    \mu = \frac{\Mstsnk}{\Delta m} \, ,
\end{equation}
where $\Delta m = 100 \, \Msun$ in our method and is meant to represent the mass at which one might expect to have at least one O-star. We will refer to the number of O-stars sampled from this distribution as $N_{\rm O, sink}$. Wind luminosity is then assigned to each O-star sampled in this way as 
\begin{equation}
    \mathcal{L}_{w,{\rm O-star}} = \Delta m\Psi_w \, .
\end{equation}
The total wind luminosity of a given sink particle is then $N_{\rm O, sink}\mathcal{L}_{w,{\rm O-star}}$.

This method was originally implemented in \citet{Su18} for galaxy scale simulations where, in general, $\Mstsnk \gg \Delta m$. Because of this, it was reasonable to assign $N_{\rm O, sink}$ in an unbiased way after one sampling from a Poisson distribution as outlined above. However, in our simulations, it is quite possible that when a star particle initially forms it has $\Mstsnk \ll \Delta m$. In our scenario, when one initially samples from the Poisson distribution it is likely that zero O-stars will be assigned. Since our sinks can grow in mass, we want to allow them \textit{eventually} to sample from the Poisson distribution again. However, if we allowed this re-sampling every time step, we would essentially be biasing the sampling process, ``flipping the coin'' over and over again until we got $N_{\rm O, sink}>0$ and then halting our sampling. To mitigate this process we sample for $N_{\rm O, sink}$ once when the star particle is formed and then afterwards only every time it has doubled in mass.

\begin{figure*}
    \centering
    \includegraphics{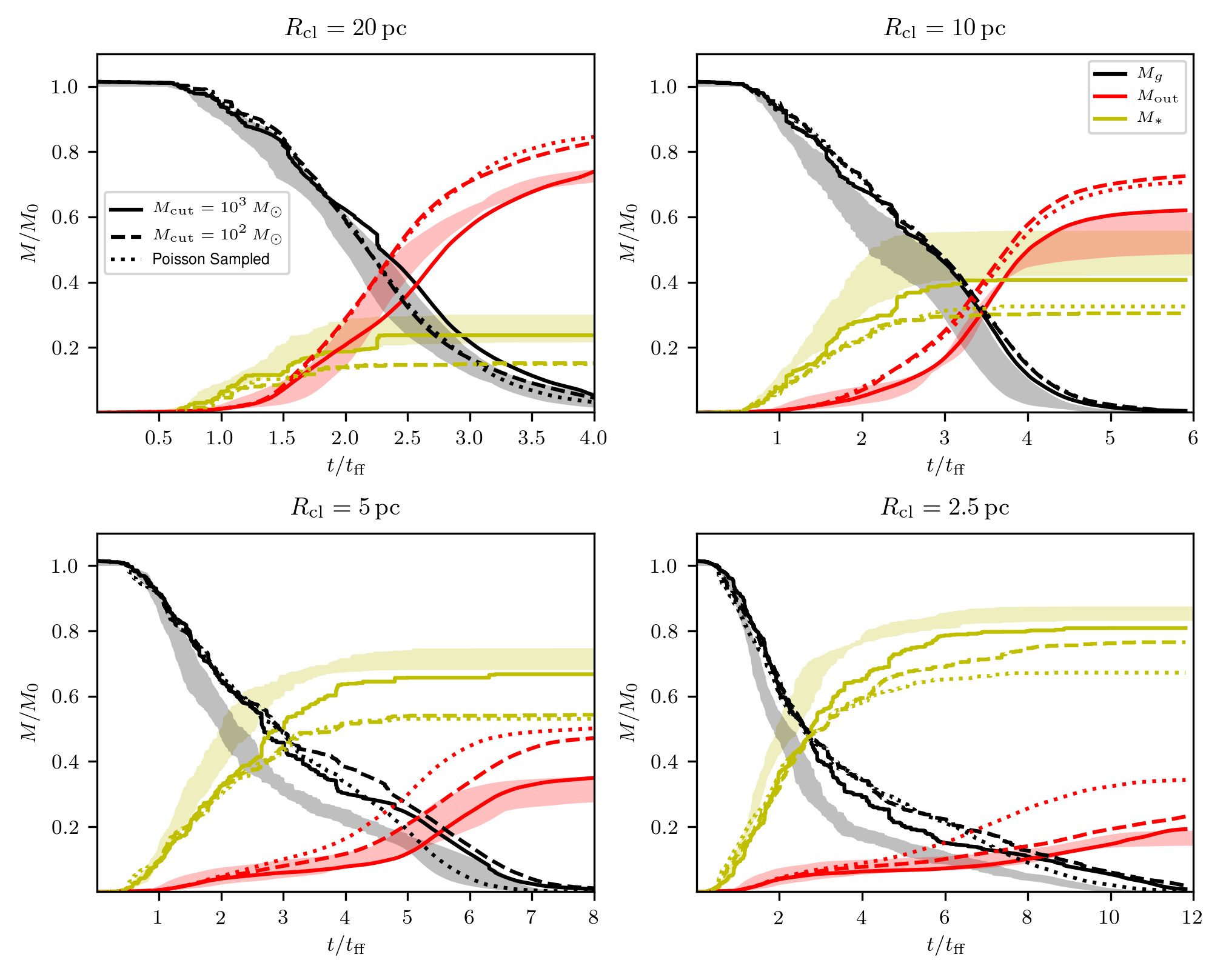}
    \caption{
    A comparison of results from different methods of assigning wind luminosity to star particles.  Line style denotes the assignment method (see text).  For each model, we show mass of gas on the grid (black), mass of stars (yellow),  and mass that has flowed out of the simulation domain (red). The shaded regions show the range of histories given by the higher resolution simulations described in the main body of the text, which differ from one another only in the random phases used to generate the initial turbulent velocity field.}
    \label{fig:model_comparison}
\end{figure*}

In \autoref{fig:model_comparison} we show a comparison of the results using the different prescriptions described above. For each model we show the history of how some gas forms stars while other gas is expelled from the cloud. In general, the models with $\Mcut = 10^2 \, \Msun$ have lower final $\sfe$ than when $\Mcut = 10^3 \, \Msun$, and result in more massive outflows earlier in the simulations. This is to be expected, since reducing $\Mcut$ to $10^2 \Msun$ from $10^3\Msun$ effectively increases the available mass that has active winds, increasing the wind luminosity and helping to drive material out of the cloud.  

Interestingly, the Poisson sampling prescription results in very similar behavior to the $\Mcut = 10^2 \, \Msun$ simulations, with the outflow and star formation histories only differing significantly for the densest simulations/smallest clouds. From inspecting the wind luminosities and distributions in $\Mstsnk$ and $N_{{\rm O, sink}}$ in these simulations we have determined that this stronger outflow is due to a larger effective wind luminosity caused by more sink particles with $\Mstsnk < \Delta m$ having $N_{\rm O, sink}\geq 1$ than expected from direct sampling. This is due to the above-mentioned sampling problem that was evidently only partially mitigated in our denser simulations by the fix we applied. This could potentially be ameliorated by a stricter sampling prescription, but exploring that possibility is beyond the scope of these tests.

As a comparison for the amount of variation between models, we also show the range of SFHs of the simulations presented in the main text which differ only in the initial turbulent velocity field (but are run at higher resolution than the simulations discussed above). These ranges are shown as shaded regions in \autoref{fig:model_comparison}. For the lowest density clouds \autoref{fig:model_comparison} shows that the spread in results for varying luminosity assignment choices is similar to the variation seen between models with different turbulence seeds, but this is not true in the higher density simulations. For this reason, the prescription for assigning feedback to sink particles remains an important factor to consider when investigating star-forming cloud simulations.

%\begin{deluxetable}{cccc}
%\tablecaption{$\Mcut$ parameter exploration.\label{tab:sim_fmcut}}
%\tablewidth{0pt}
%\tablehead{Cloud Radius & $\Lbox/\Delta x$ & $\Mcut$ & $f_{>\Mcut, {\rm final}}\ ^a$ \\
%$[{\rm pc}]$ &  & $[\Msun]$ & }
%\startdata
%20   & 128 & $10^2$ & 1.0   \\
%20  & 128 & $10^3$ & 0.76  \\
%20  & 256 & $10^3$ & 0.78  \\ \hline
%10   & 128 & $10^2$ & 0.99  \\
%10  & 128 & $10^3$ & 0.86  \\
%10  & 256 & $10^3$ & 0.82  \\ \hline
%5    & 128 & $10^2$ & 0.98  \\
%5   & 128 & $10^3$ & 0.99  \\
%5   & 256 & $10^3$ & 0.94  \\ \hline
%2.5  & 128 & $10^2$ & 0.99  \\
%2.5 & 128 & $10^3$ & 0.99  \\
%2.5 & 256 & $10^3$ & 0.98  \\
%\enddata
%%\tablecomments{Each}
%\tablenotetext{a}{Fraction of stellar mass above $\Mcut$ at the end of the simulation.}
%\end{deluxetable}

\begin{figure*}
    \centering
    \includegraphics{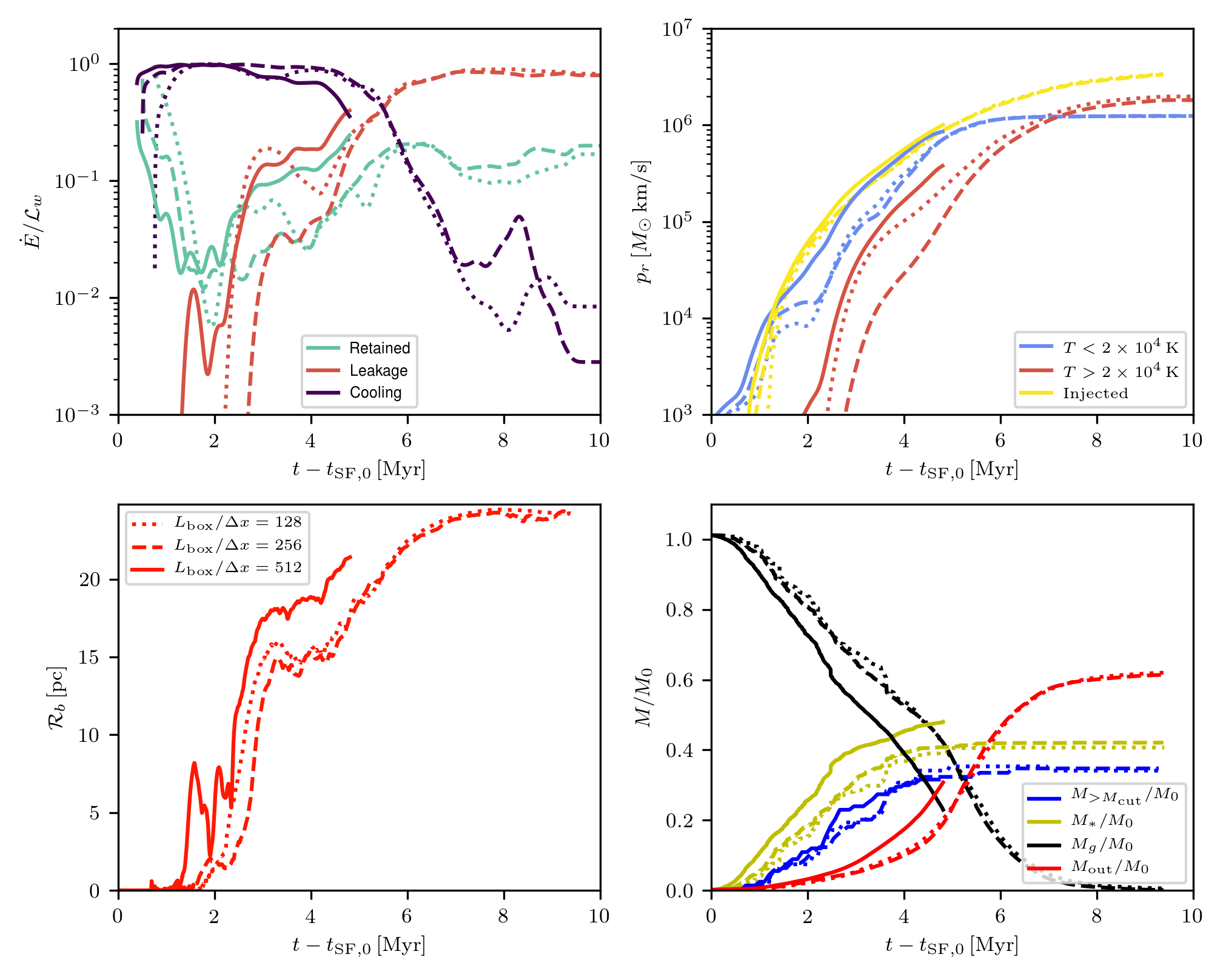}
    \caption{A resolution study of key history parameters for model \textbf{R10}. The line style denotes results from simulations with low resolution (dotted), fiducial resolution (dashed), and high resolution (solid). \textit{Upper left}: wind energy losses and retained fraction (see labels), as in left column of \autoref{fig:energy_loss}. \textit{Upper right}: radial momentum carried in cool/warm gas and hot gas, along with the net momentum injected by the wind, as in middle column of \autoref{fig:energy_loss}. \textit{Lower left}: effective radius of the wind driven bubble, as in the right column of \autoref{fig:energy_loss}. \textit{Lower right}: mass history, showing total gas mass on the grid, gas advected out of the domain, and stellar mass, as in \autoref{fig:model_comparison}.  Also shown is the mass of stars above $\Mcut = 10^3 \, \Msun$, which produces winds.}
    \label{fig:resolution}
\end{figure*}

\section{Resolution Study}
\label{app:resolution}

Here, we provide a resolution study of our simulations for the {\bf R10} cloud model. The simulations presented here use a single set of random phases to initialize the turbulent velocity spectrum and all of the same numerical techniques as described in \autoref{sec:methods}. Here we use the same prescription for assigning wind feedback to sink particles as described in the text, with $\Mcut = 10^3\, \Msun$. All simulations have the same parameters as outlined in row 2 and columns 2-5 of \autoref{tab:sim_params}. The only parameter varied is the spatial resolution, $\Delta x$, given in column 6 of \autoref{tab:sim_params}. Specifically, we perform a simulation at half the nominal resolution given in \autoref{tab:sim_params} and a simulation at twice the nominal resolution, but for about half of the full simulation time due to computational cost.

\autoref{fig:resolution} presents evolution of several key quantities across the different resolutions. One quantity which we have not shown elsewhere is $M_{>\Mcut}$ which we define as the total mass of star particles that are more massive than $\Mcut$, the threshold used to determine whether or not a star particle has a wind. The total stellar mass and $\Psi_w$ determine the total wind luminosity (modulo the effects of independently aging star particles, which causes $\Psi_w$ to vary mildly from source to source).

The simulations run at low and fiducial resolution (dotted and dashed lines) are quite similar: they have comparable energy loss histories and corresponding radial momentum ($t_{\rm pr}$) and breakout ($t_{\rm br}$) times, the evolution of their bubble sizes track one another, and their star formation histories are nearly identical. However, there is a very slight increase in the total stellar mass formed going from the low to the fiducial resolution. For the high resolution model, this trend continues and is greater in magnitude, with both a higher star formation rate and higher lifetime SFE than in the fiducial and low-resolution models. In contrast, the histories of the mass in particles that produce winds, $M_{>\Mcut}$ (blue lines in bottom right panel), is very similar across all resolutions. This is also mirrored by the similarity in the equivalent injected total wind momentum  (yellow lines in upper right panel of \autoref{fig:resolution}).  

The differences/similarities in $M_*$ vs. $M_{>\Mcut}$ histories reflect both the reduction in the mean initial sink particle mass at higher resolution, based on the particle creation criteria (see also \citet{Grudic21} or \citet{Haugbolle18} for a similar effect), and the prescription for assigning winds to sink particles.  This comparison also provides additional evidence of the importance of feedback to limiting star formation. In the highest resolution simulations, many low-mass sink particles form that never grow sufficiently to have winds associated with them, and therefore do not help to disperse high-density collapsing structures or lower-density surrounding material.  Rather, star formation continues until the mass in particles that \textit{do} have winds is sufficient to remove the remaining cloud material. For the fiducial and low-resolution models, at late stages only $\sim 15-20\%$ of the stellar mass is in sink particles that do not provide feedback, but this increases to $\sim 30\%$ in the high-resolution model.  This high-resolution  model is therefore significantly ``underpowered'' relative to what it should be for a fully sampled IMF, accounting for its enhanced SFE.  

We note that the highest resolution simulation transitions more quickly to the ``breakout'' phase than the fiducial and low-resolution models, based on the earlier dominance of leakage losses over cooling losses in the upper left panel of \autoref{fig:resolution}. This could be due to some combination of the higher-resolution model having less ambient gas due to more total star formation (lower-right of \autoref{fig:resolution}) or higher contrasts in the density at high resolution, which increases porosity. The more rapid transition to breakout is also reflected in the more rapid growth of ${\cal R}_b$ for the highest resolution model (lower-left panel of \autoref{fig:resolution}).   

The differences in results as a function of resolution appear to arise from imposition of a mass cut that is far above the minimum sink mass at high resolution, so that many low mass sinks are ``left behind'' and never accrete enough ambient gas to reach $\Mcut$ and generate winds.  In contrast, for the fiducial and low resolution models, inspection of late-time mass functions show that almost all particles have grown enough to host winds, consistent with the original design of our wind assignment prescription employing $\Mcut$.  In future work, it will clearly be important to design a wind-assignment prescription that is more robust to changes in the sink mass function with resolution, while still having an astronomically realistic and numerically practical number of feedback sources.  One way to do this is by directly sampling from a mass function to assign individual massive stars (with associated feedback) to the sink particles \citep[e.g.][]{Sormani2017}.  

\bibliography{bibliography}{}

\begin{thebibliography}{}
\expandafter\ifx\csname natexlab\endcsname\relax\def\natexlab#1{#1}\fi
\providecommand{\url}[1]{\href{#1}{#1}}
\providecommand{\dodoi}[1]{doi:~\href{http://doi.org/#1}{\nolinkurl{#1}}}
\providecommand{\doeprint}[1]{\href{http://ascl.net/#1}{\nolinkurl{http://ascl.net/#1}}}
\providecommand{\doarXiv}[1]{\href{https://arxiv.org/abs/#1}{\nolinkurl{https://arxiv.org/abs/#1}}}

\bibitem[{{Agertz} {et~al.}(2013){Agertz}, {Kravtsov}, {Leitner}, \&
  {Gnedin}}]{Agertz_2013}
{Agertz}, O., {Kravtsov}, A.~V., {Leitner}, S.~N., \& {Gnedin}, N.~Y. 2013,
  \apj, 770, 25, \dodoi{10.1088/0004-637X/770/1/25}

\bibitem[{{Astropy Collaboration} {et~al.}(2013){Astropy Collaboration},
  {Robitaille}, {Tollerud}, {Greenfield}, {Droettboom}, {Bray}, {Aldcroft},
  {Davis}, {Ginsburg}, {Price-Whelan}, {Kerzendorf}, {Conley}, {Crighton},
  {Barbary}, {Muna}, {Ferguson}, {Grollier}, {Parikh}, {Nair}, {Unther},
  {Deil}, {Woillez}, {Conseil}, {Kramer}, {Turner}, {Singer}, {Fox}, {Weaver},
  {Zabalza}, {Edwards}, {Azalee Bostroem}, {Burke}, {Casey}, {Crawford},
  {Dencheva}, {Ely}, {Jenness}, {Labrie}, {Lim}, {Pierfederici}, {Pontzen},
  {Ptak}, {Refsdal}, {Servillat}, \& {Streicher}}]{astropy13}
{Astropy Collaboration}, {Robitaille}, T.~P., {Tollerud}, E.~J., {et~al.} 2013,
  \aap, 558, A33, \dodoi{10.1051/0004-6361/201322068}

\bibitem[{{Astropy Collaboration} {et~al.}(2018){Astropy Collaboration},
  {Price-Whelan}, {Sip{\H{o}}cz}, {G{\"u}nther}, {Lim}, {Crawford}, {Conseil},
  {Shupe}, {Craig}, {Dencheva}, {Ginsburg}, {Vand erPlas}, {Bradley},
  {P{\'e}rez-Su{\'a}rez}, {de Val-Borro}, {Aldcroft}, {Cruz}, {Robitaille},
  {Tollerud}, {Ardelean}, {Babej}, {Bach}, {Bachetti}, {Bakanov}, {Bamford},
  {Barentsen}, {Barmby}, {Baumbach}, {Berry}, {Biscani}, {Boquien}, {Bostroem},
  {Bouma}, {Brammer}, {Bray}, {Breytenbach}, {Buddelmeijer}, {Burke},
  {Calderone}, {Cano Rodr{\'\i}guez}, {Cara}, {Cardoso}, {Cheedella}, {Copin},
  {Corrales}, {Crichton}, {D'Avella}, {Deil}, {Depagne}, {Dietrich}, {Donath},
  {Droettboom}, {Earl}, {Erben}, {Fabbro}, {Ferreira}, {Finethy}, {Fox},
  {Garrison}, {Gibbons}, {Goldstein}, {Gommers}, {Greco}, {Greenfield},
  {Groener}, {Grollier}, {Hagen}, {Hirst}, {Homeier}, {Horton}, {Hosseinzadeh},
  {Hu}, {Hunkeler}, {Ivezi{\'c}}, {Jain}, {Jenness}, {Kanarek}, {Kendrew},
  {Kern}, {Kerzendorf}, {Khvalko}, {King}, {Kirkby}, {Kulkarni}, {Kumar},
  {Lee}, {Lenz}, {Littlefair}, {Ma}, {Macleod}, {Mastropietro}, {McCully},
  {Montagnac}, {Morris}, {Mueller}, {Mumford}, {Muna}, {Murphy}, {Nelson},
  {Nguyen}, {Ninan}, {N{\"o}the}, {Ogaz}, {Oh}, {Parejko}, {Parley}, {Pascual},
  {Patil}, {Patil}, {Plunkett}, {Prochaska}, {Rastogi}, {Reddy Janga},
  {Sabater}, {Sakurikar}, {Seifert}, {Sherbert}, {Sherwood-Taylor}, {Shih},
  {Sick}, {Silbiger}, {Singanamalla}, {Singer}, {Sladen}, {Sooley},
  {Sornarajah}, {Streicher}, {Teuben}, {Thomas}, {Tremblay}, {Turner},
  {Terr{\'o}n}, {van Kerkwijk}, {de la Vega}, {Watkins}, {Weaver}, {Whitmore},
  {Woillez}, {Zabalza}, \& {Astropy Contributors}}]{astropy18}
{Astropy Collaboration}, {Price-Whelan}, A.~M., {Sip{\H{o}}cz}, B.~M., {et~al.}
  2018, \aj, 156, 123, \dodoi{10.3847/1538-3881/aabc4f}

\bibitem[{{Bailin}(2018)}]{Bailin18}
{Bailin}, J. 2018, \apj, 863, 99, \dodoi{10.3847/1538-4357/aad178}

\bibitem[{{Bastian} \& {Lardo}(2018)}]{BastianLardo18}
{Bastian}, N., \& {Lardo}, C. 2018, \araa, 56, 83,
  \dodoi{10.1146/annurev-astro-081817-051839}

\bibitem[{{Cheng} {et~al.}(2021){Cheng}, {Wang}, \& {Lim}}]{Cheng21}
{Cheng}, Y., {Wang}, Q.~D., \& {Lim}, S. 2021, \mnras, 504, 1627,
  \dodoi{10.1093/mnras/stab1040}

\bibitem[{{Chevance} {et~al.}(2020){Chevance}, {Kruijssen}, {Vazquez-Semadeni},
  {Nakamura}, {Klessen}, {Ballesteros-Paredes}, {Inutsuka}, {Adamo}, \&
  {Hennebelle}}]{ChevanceRev20}
{Chevance}, M., {Kruijssen}, J.~M.~D., {Vazquez-Semadeni}, E., {et~al.} 2020,
  \ssr, 216, 50, \dodoi{10.1007/s11214-020-00674-x}

\bibitem[{{Choi} {et~al.}(2016){Choi}, {Dotter}, {Conroy}, {Cantiello},
  {Paxton}, \& {Johnson}}]{MIST}
{Choi}, J., {Dotter}, A., {Conroy}, C., {et~al.} 2016, \apj, 823, 102,
  \dodoi{10.3847/0004-637X/823/2/102}

\bibitem[{{Dale} {et~al.}(2012){Dale}, {Ercolano}, \& {Bonnell}}]{Dale12}
{Dale}, J.~E., {Ercolano}, B., \& {Bonnell}, I.~A. 2012, \mnras, 424, 377,
  \dodoi{10.1111/j.1365-2966.2012.21205.x}

\bibitem[{{Dale} {et~al.}(2013){Dale}, {Ercolano}, \& {Bonnell}}]{Dale_2013a}
---. 2013, \mnras, 430, 234, \dodoi{10.1093/mnras/sts592}

\bibitem[{{Dale} {et~al.}(2014){Dale}, {Ngoumou}, {Ercolano}, \&
  {Bonnell}}]{Dale14}
{Dale}, J.~E., {Ngoumou}, J., {Ercolano}, B., \& {Bonnell}, I.~A. 2014, \mnras,
  442, 694, \dodoi{10.1093/mnras/stu816}

\bibitem[{{Ekstr{\"o}m} {et~al.}(2012){Ekstr{\"o}m}, {Georgy}, {Eggenberger},
  {Meynet}, {Mowlavi}, {Wyttenbach}, {Granada}, {Decressin}, {Hirschi},
  {Frischknecht}, {Charbonnel}, \& {Maeder}}]{Ekstrom12}
{Ekstr{\"o}m}, S., {Georgy}, C., {Eggenberger}, P., {et~al.} 2012, \aap, 537,
  A146, \dodoi{10.1051/0004-6361/201117751}

\bibitem[{{Emig} {et~al.}(2020){Emig}, {Bolatto}, {Leroy}, {Mills}, {Jimenez
  Donaire}, {Tielens}, {Ginsburg}, {Gorski}, {Krieger}, {Levy}, {Meier}, {Ott},
  {Rosolowsky}, {Thompson}, \& {Veilleux}}]{Emig20}
{Emig}, K.~L., {Bolatto}, A.~D., {Leroy}, A.~K., {et~al.} 2020, arXiv e-prints,
  arXiv:2009.05154.
\newblock \doarXiv{2009.05154}

\bibitem[{{Evans} {et~al.}(2021){Evans}, {Heyer. Marc-Antoine
  Miville-Desch{\^e}nes}, \& {Merello}}]{Evans21}
{Evans}, Neal~J., I., {Heyer. Marc-Antoine Miville-Desch{\^e}nes}, M., \&
  {Merello}, Q. N.-L.~M. 2021, arXiv e-prints, arXiv:2107.05750.
\newblock \doarXiv{2107.05750}

\bibitem[{{Fall} {et~al.}(2010){Fall}, {Krumholz}, \& {Matzner}}]{Fall10}
{Fall}, S.~M., {Krumholz}, M.~R., \& {Matzner}, C.~D. 2010, \apjl, 710, L142,
  \dodoi{10.1088/2041-8205/710/2/L142}

\bibitem[{{Freeman} \& {Bland-Hawthorn}(2002)}]{FBH02}
{Freeman}, K., \& {Bland-Hawthorn}, J. 2002, \araa, 40, 487,
  \dodoi{10.1146/annurev.astro.40.060401.093840}

\bibitem[{{Fujita} {et~al.}(2021){Fujita}, {Tsutsumi}, {Ohama}, {Habe},
  {Sakre}, {Okawa}, {Kohno}, {Hattori}, {Nishimura}, {Torii}, {Sano},
  {Tachihara}, {Kimura}, {Ogawa}, \& {Fukui}}]{FujitaOrion21}
{Fujita}, S., {Tsutsumi}, D., {Ohama}, A., {et~al.} 2021, \pasj, 73, S273,
  \dodoi{10.1093/pasj/psaa005}

\bibitem[{{Geen} {et~al.}(2021){Geen}, {Bieri}, {Rosdahl}, \& {de
  Koter}}]{Geen21}
{Geen}, S., {Bieri}, R., {Rosdahl}, J., \& {de Koter}, A. 2021, \mnras, 501,
  1352, \dodoi{10.1093/mnras/staa3705}

\bibitem[{{Geen} {et~al.}(2016){Geen}, {Hennebelle}, {Tremblin}, \&
  {Rosdahl}}]{Geen_2016}
{Geen}, S., {Hennebelle}, P., {Tremblin}, P., \& {Rosdahl}, J. 2016, \mnras,
  463, 3129, \dodoi{10.1093/mnras/stw2235}

\bibitem[{{Gong} \& {Ostriker}(2013)}]{GongOstriker13}
{Gong}, H., \& {Ostriker}, E.~C. 2013, \apjs, 204, 8,
  \dodoi{10.1088/0067-0049/204/1/8}

\bibitem[{{Gratton} {et~al.}(2019){Gratton}, {Bragaglia}, {Carretta},
  {D'Orazi}, {Lucatello}, \& {Sollima}}]{Gratton19}
{Gratton}, R., {Bragaglia}, A., {Carretta}, E., {et~al.} 2019, \aapr, 27, 8,
  \dodoi{10.1007/s00159-019-0119-3}

\bibitem[{{Grudi{\'c}} {et~al.}(2021){Grudi{\'c}}, {Guszejnov}, {Hopkins},
  {Offner}, \& {Faucher-Gigu{\'e}re}}]{Grudic21}
{Grudi{\'c}}, M.~Y., {Guszejnov}, D., {Hopkins}, P.~F., {Offner}, S. S.~R., \&
  {Faucher-Gigu{\'e}re}, C.-A. 2021, \mnras, \dodoi{10.1093/mnras/stab1347}

\bibitem[{{Grudi{\'c}} \& {Hopkins}(2019)}]{Grudic19}
{Grudi{\'c}}, M.~Y., \& {Hopkins}, P.~F. 2019, \mnras, 488, 2970,
  \dodoi{10.1093/mnras/stz1820}

\bibitem[{{Harper-Clark} \& {Murray}(2009)}]{HCM09}
{Harper-Clark}, E., \& {Murray}, N. 2009, \apj, 693, 1696,
  \dodoi{10.1088/0004-637X/693/2/1696}

\bibitem[{{Harris} {et~al.}(2020){Harris}, {Jarrod Millman}, {van der Walt},
  {Gommers}, {Virtanen}, {Cournapeau}, {Wieser}, {Taylor}, {Berg}, {Smith},
  {Kern}, {Picus}, {Hoyer}, {van Kerkwijk}, {Brett}, {Haldane}, {Fern{\'a}ndez
  del R{\'\i}o}, {Wiebe}, {Peterson}, {G{\'e}rard-Marchant}, {Sheppard},
  {Reddy}, {Weckesser}, {Abbasi}, {Gohlke}, \& {Oliphant}}]{harrisNumpy2020}
{Harris}, C.~R., {Jarrod Millman}, K., {van der Walt}, S.~J., {et~al.} 2020,
  arXiv e-prints, arXiv:2006.10256.
\newblock \doarXiv{2006.10256}

\bibitem[{{Haugb{\o}lle} {et~al.}(2018){Haugb{\o}lle}, {Padoan}, \&
  {Nordlund}}]{Haugbolle18}
{Haugb{\o}lle}, T., {Padoan}, P., \& {Nordlund}, {\r{A}}. 2018, \apj, 854, 35,
  \dodoi{10.3847/1538-4357/aaa432}

\bibitem[{{Heyer} \& {Dame}(2015)}]{HeyerDame15}
{Heyer}, M., \& {Dame}, T.~M. 2015, \araa, 53, 583,
  \dodoi{10.1146/annurev-astro-082214-122324}

\bibitem[{{Hosek} {et~al.}(2019){Hosek}, {Lu}, {Anderson}, {Najarro}, {Ghez},
  {Morris}, {Clarkson}, \& {Albers}}]{Hosek19}
{Hosek}, Matthew~W., J., {Lu}, J.~R., {Anderson}, J., {et~al.} 2019, \apj, 870,
  44, \dodoi{10.3847/1538-4357/aaef90}

\bibitem[{{Howard} {et~al.}(2017){Howard}, {Pudritz}, \& {Harris}}]{Howard17}
{Howard}, C.~S., {Pudritz}, R.~E., \& {Harris}, W.~E. 2017, \mnras, 470, 3346,
  \dodoi{10.1093/mnras/stx1363}

\bibitem[{{Hoyer} {et~al.}(2017){Hoyer}, {Hamman}, {Fitzgerald}, {Kleeman},
  {Maussion}, {Kluyver}, {Munroe}, {Roos}, {Fujii}, {Wolfram}, {Markel},
  {Helmus}, {Bovy}, {Hatfield Dodds}, {crusaderky}, {Cable}, {Abernathey},
  {Bell}, {Noel}, {Kanmae}, {Miles}, {Hill}, {chunweiyuan}, {Sinclair},
  {Signell}, {Oleksandr}, {Filipe}, {Lalibert{\'e}}, {Rothenberg}, \&
  {Malevich}}]{hoyer17}
{Hoyer}, S., {Hamman}, J., {Fitzgerald}, C., {et~al.} 2017, {Pydata/Xarray:
  V0.10.0}, v0.10.0,  Zenodo, \dodoi{10.5281/zenodo.1063607}

\bibitem[{{Hunter}(2007)}]{matplotlib_hunter07}
{Hunter}, J.~D. 2007, Computing in Science and Engineering, 9, 90,
  \dodoi{10.1109/MCSE.2007.55}

\bibitem[{{Iffrig} \& {Hennebelle}(2015)}]{Iffrig_Hennebelle2015}
{Iffrig}, O., \& {Hennebelle}, P. 2015, \aap, 576, A95,
  \dodoi{10.1051/0004-6361/201424556}

\bibitem[{{Johnson} {et~al.}(2015){Johnson}, {Leroy}, {Indebetouw}, {Brogan},
  {Whitmore}, {Hibbard}, {Sheth}, \& {Evans}}]{Johnson15}
{Johnson}, K.~E., {Leroy}, A.~K., {Indebetouw}, R., {et~al.} 2015, \apj, 806,
  35, \dodoi{10.1088/0004-637X/806/1/35}

\bibitem[{{Kim} {et~al.}(2008){Kim}, {Kim}, \& {Ostriker}}]{Kim08}
{Kim}, C.-G., {Kim}, W.-T., \& {Ostriker}, E.~C. 2008, \apj, 681, 1148,
  \dodoi{10.1086/588752}

\bibitem[{{Kim} \& {Ostriker}(2017)}]{CGK_TIGRESS1}
{Kim}, C.-G., \& {Ostriker}, E.~C. 2017, \apj, 846, 133,
  \dodoi{10.3847/1538-4357/aa8599}

\bibitem[{{Kim} {et~al.}(2020){Kim}, {Ostriker}, {Somerville}, {Bryan},
  {Fielding}, {Forbes}, {Hayward}, {Hernquist}, \&
  {Pandya}}]{Kim_Ostriker_SMAUG2020}
{Kim}, C.-G., {Ostriker}, E.~C., {Somerville}, R.~S., {et~al.} 2020, \apj, 900,
  61, \dodoi{10.3847/1538-4357/aba962}

\bibitem[{{Kim} {et~al.}(2016){Kim}, {Kim}, \& {Ostriker}}]{Kim_JG2016}
{Kim}, J.-G., {Kim}, W.-T., \& {Ostriker}, E.~C. 2016, \apj, 819, 137,
  \dodoi{10.3847/0004-637X/819/2/137}

\bibitem[{{Kim} {et~al.}(2018){Kim}, {Kim}, \& {Ostriker}}]{JGK18}
---. 2018, \apj, 859, 68, \dodoi{10.3847/1538-4357/aabe27}

\bibitem[{{Kim} {et~al.}(2019){Kim}, {Kim}, \& {Ostriker}}]{JGK19}
---. 2019, \apj, 883, 102, \dodoi{10.3847/1538-4357/ab3d3d}

\bibitem[{{Kim} {et~al.}(2021){Kim}, {Ostriker}, \& {Filippova}}]{KOF_2021}
{Kim}, J.-G., {Ostriker}, E.~C., \& {Filippova}, N. 2021, \apj, 911, 128,
  \dodoi{10.3847/1538-4357/abe934}

\bibitem[{{Koyama} \& {Inutsuka}(2002)}]{KoyamaInutsuka02}
{Koyama}, H., \& {Inutsuka}, S.-i. 2002, \apjl, 564, L97,
  \dodoi{10.1086/338978}

\bibitem[{{Kroupa}(2001)}]{KroupaIMF}
{Kroupa}, P. 2001, \mnras, 322, 231, \dodoi{10.1046/j.1365-8711.2001.04022.x}

\bibitem[{{Krumholz} \& {Matzner}(2009)}]{Krumholz_Matzner2009}
{Krumholz}, M.~R., \& {Matzner}, C.~D. 2009, \apj, 703, 1352,
  \dodoi{10.1088/0004-637X/703/2/1352}

\bibitem[{{Krumholz} {et~al.}(2019){Krumholz}, {McKee}, \& {Bland
  -Hawthorn}}]{KMBH19}
{Krumholz}, M.~R., {McKee}, C.~F., \& {Bland -Hawthorn}, J. 2019, \araa, 57,
  227, \dodoi{10.1146/annurev-astro-091918-104430}

\bibitem[{{Lancaster} {et~al.}(2021{\natexlab{a}}){Lancaster}, {Ostriker},
  {Kim}, \& {Kim}}]{Lancaster21a}
{Lancaster}, L., {Ostriker}, E.~C., {Kim}, J.-G., \& {Kim}, C.-G.
  2021{\natexlab{a}}, \apj, 914, 89, \dodoi{10.3847/1538-4357/abf8ab}

\bibitem[{{Lancaster} {et~al.}(2021{\natexlab{b}}){Lancaster}, {Ostriker},
  {Kim}, \& {Kim}}]{Lancaster21b}
---. 2021{\natexlab{b}}, \apj, 914, 90, \dodoi{10.3847/1538-4357/abf8ac}

\bibitem[{{Larson}(1969)}]{Larson69}
{Larson}, R.~B. 1969, \mnras, 145, 405, \dodoi{10.1093/mnras/145.4.405}

\bibitem[{{Leitherer} {et~al.}(2014){Leitherer}, {Ekstr{\"o}m}, {Meynet},
  {Schaerer}, {Agienko}, \& {Levesque}}]{Leitherer14}
{Leitherer}, C., {Ekstr{\"o}m}, S., {Meynet}, G., {et~al.} 2014, \apjs, 212,
  14, \dodoi{10.1088/0067-0049/212/1/14}

\bibitem[{{Leitherer} {et~al.}(1992){Leitherer}, {Robert}, \&
  {Drissen}}]{Leitherer92}
{Leitherer}, C., {Robert}, C., \& {Drissen}, L. 1992, \apj, 401, 596,
  \dodoi{10.1086/172089}

\bibitem[{{Leitherer} {et~al.}(1999){Leitherer}, {Schaerer}, {Goldader},
  {Delgado}, {Robert}, {Kune}, {de Mello}, {Devost}, \& {Heckman}}]{SB99}
{Leitherer}, C., {Schaerer}, D., {Goldader}, J.~D., {et~al.} 1999, \apjs, 123,
  3, \dodoi{10.1086/313233}

\bibitem[{{Leroy} {et~al.}(2018){Leroy}, {Bolatto}, {Ostriker}, {Walter},
  {Gorski}, {Ginsburg}, {Krieger}, {Levy}, {Meier}, {Mills}, {Ott},
  {Rosolowsky}, {Thompson}, {Veilleux}, \& {Zschaechner}}]{Leroy18}
{Leroy}, A.~K., {Bolatto}, A.~D., {Ostriker}, E.~C., {et~al.} 2018, \apj, 869,
  126, \dodoi{10.3847/1538-4357/aaecd1}

\bibitem[{{Levy} {et~al.}(2021){Levy}, {Bolatto}, {Leroy}, {Emig}, {Gorski},
  {Krieger}, {Lenki{\'c}}, {Meier}, {Mills}, {Ott}, {Rosolowsky}, {Tarantino},
  {Veilleux}, {Walter}, {Wei{\ss}}, \& {Zwaan}}]{Levy20}
{Levy}, R.~C., {Bolatto}, A.~D., {Leroy}, A.~K., {et~al.} 2021, \apj, 912, 4,
  \dodoi{10.3847/1538-4357/abec84}

\bibitem[{{Li} {et~al.}(2019){Li}, {Vogelsberger}, {Marinacci}, \&
  {Gnedin}}]{Li_Vogelsberger2019}
{Li}, H., {Vogelsberger}, M., {Marinacci}, F., \& {Gnedin}, O.~Y. 2019, \mnras,
  487, 364, \dodoi{10.1093/mnras/stz1271}

\bibitem[{{Lochhaas} \& {Thompson}(2017)}]{LochhaasThompson17}
{Lochhaas}, C., \& {Thompson}, T.~A. 2017, \mnras, 470, 977,
  \dodoi{10.1093/mnras/stx1289}

\bibitem[{{Lopez} {et~al.}(2011){Lopez}, {Krumholz}, {Bolatto}, {Prochaska}, \&
  {Ramirez-Ruiz}}]{Lopez11}
{Lopez}, L.~A., {Krumholz}, M.~R., {Bolatto}, A.~D., {Prochaska}, J.~X., \&
  {Ramirez-Ruiz}, E. 2011, \apj, 731, 91, \dodoi{10.1088/0004-637X/731/2/91}

\bibitem[{{Lopez} {et~al.}(2014){Lopez}, {Krumholz}, {Bolatto}, {Prochaska},
  {Ramirez-Ruiz}, \& {Castro}}]{Lopez14}
{Lopez}, L.~A., {Krumholz}, M.~R., {Bolatto}, A.~D., {et~al.} 2014, \apj, 795,
  121, \dodoi{10.1088/0004-637X/795/2/121}

\bibitem[{{Lu} {et~al.}(2013){Lu}, {Do}, {Ghez}, {Morris}, {Yelda}, \&
  {Matthews}}]{Lu13}
{Lu}, J.~R., {Do}, T., {Ghez}, A.~M., {et~al.} 2013, \apj, 764, 155,
  \dodoi{10.1088/0004-637X/764/2/155}

\bibitem[{{McCrady} {et~al.}(2005){McCrady}, {Graham}, \& {Vacca}}]{McCrady05}
{McCrady}, N., {Graham}, J.~R., \& {Vacca}, W.~D. 2005, \apj, 621, 278,
  \dodoi{10.1086/427487}

\bibitem[{{Murray}(2011)}]{Murray11}
{Murray}, N. 2011, \apj, 729, 133, \dodoi{10.1088/0004-637X/729/2/133}

\bibitem[{{Oey} {et~al.}(2017){Oey}, {Herrera}, {Silich}, {Reiter}, {James},
  {Jaskot}, \& {Micheva}}]{Oey17}
{Oey}, M.~S., {Herrera}, C.~N., {Silich}, S., {et~al.} 2017, \apjl, 849, L1,
  \dodoi{10.3847/2041-8213/aa9215}

\bibitem[{{Penston}(1969)}]{Penston69}
{Penston}, M.~V. 1969, \mnras, 144, 425, \dodoi{10.1093/mnras/144.4.425}

\bibitem[{{Perez} \& {Granger}(2007)}]{Perez07}
{Perez}, F., \& {Granger}, B.~E. 2007, Computing in Science and Engineering, 9,
  21, \dodoi{10.1109/MCSE.2007.53}

\bibitem[{{Raskutti} {et~al.}(2016){Raskutti}, {Ostriker}, \&
  {Skinner}}]{Raskutti16}
{Raskutti}, S., {Ostriker}, E.~C., \& {Skinner}, M.~A. 2016, \apj, 829, 130,
  \dodoi{10.3847/0004-637X/829/2/130}

\bibitem[{{Raskutti} {et~al.}(2017){Raskutti}, {Ostriker}, \&
  {Skinner}}]{Raskutti17}
---. 2017, \apj, 850, 112, \dodoi{10.3847/1538-4357/aa965e}

\bibitem[{{Reback} {et~al.}(2020){Reback}, {McKinney}, {Jbrockmendel}, {Den Van
  Bossche}, {Augspurger}, {Cloud}, {Gfyoung}, {Sinhrks}, {Klein}, {Roeschke},
  {Hawkins}, {Tratner}, {She}, {Ayd}, {Petersen}, {Garcia}, {Schendel},
  {Hayden}, {MomIsBestFriend}, {Jancauskas}, {Battiston}, {Seabold},
  {Chris-B1}, {H-Vetinari}, {Hoyer}, {Overmeire}, {Alimcmaster1}, {Dong},
  {Whelan}, \& {Mehyar}}]{pandas2020}
{Reback}, J., {McKinney}, W., {Jbrockmendel}, {et~al.} 2020,
  {pandas-dev/pandas: Pandas 1.0.3}, v1.0.3,  Zenodo,
  \dodoi{10.5281/zenodo.3509134}

\bibitem[{{Ressler} {et~al.}(2020){Ressler}, {Quataert}, \&
  {Stone}}]{Ressler20}
{Ressler}, S.~M., {Quataert}, E., \& {Stone}, J.~M. 2020, \mnras, 492, 3272,
  \dodoi{10.1093/mnras/stz3605}

\bibitem[{{Roe}(1981)}]{Roe81}
{Roe}, P.~L. 1981, Journal of Computational Physics, 43, 357,
  \dodoi{10.1016/0021-9991(81)90128-5}

\bibitem[{{Rogers} \& {Pittard}(2013)}]{Rogers_Pittard2013}
{Rogers}, H., \& {Pittard}, J.~M. 2013, \mnras, 431, 1337,
  \dodoi{10.1093/mnras/stt255}

\bibitem[{{Rosen} {et~al.}(2014){Rosen}, {Lopez}, {Krumholz}, \&
  {Ramirez-Ruiz}}]{Rosen14}
{Rosen}, A.~L., {Lopez}, L.~A., {Krumholz}, M.~R., \& {Ramirez-Ruiz}, E. 2014,
  \mnras, 442, 2701, \dodoi{10.1093/mnras/stu1037}

\bibitem[{{Schneider} {et~al.}(2020){Schneider}, {Simon}, {Guevara},
  {Buchbender}, {Higgins}, {Okada}, {Stutzki}, {G{\"u}sten}, {Anderson},
  {Bally}, {Beuther}, {Bonne}, {Bontemps}, {Chambers}, {Csengeri}, {Graf},
  {Gusdorf}, {Jacobs}, {Justen}, {Kabanovic}, {Karim}, {Luisi}, {Menten},
  {Mertens}, {Mookerjea}, {Ossenkopf-Okada}, {Pabst}, {Pound}, {Richter},
  {Reyes}, {Ricken}, {R{\"o}llig}, {Russeil}, {S{\'a}nchez-Monge}, {Sandell},
  {Tiwari}, {Wiesemeyer}, {Wolfire}, {Wyrowski}, {Zavagno}, \&
  {Tielens}}]{SOFIAFeedback20}
{Schneider}, N., {Simon}, R., {Guevara}, C., {et~al.} 2020, \pasp, 132, 104301,
  \dodoi{10.1088/1538-3873/aba840}

\bibitem[{{Skinner} \& {Ostriker}(2015)}]{Skinner_Ostriker2015}
{Skinner}, M.~A., \& {Ostriker}, E.~C. 2015, \apj, 809, 187,
  \dodoi{10.1088/0004-637X/809/2/187}

\bibitem[{{Smith}(2021)}]{Smith21}
{Smith}, M.~C. 2021, \mnras, 502, 5417, \dodoi{10.1093/mnras/stab291}

\bibitem[{{Smith}(2014)}]{Smith14}
{Smith}, N. 2014, \araa, 52, 487, \dodoi{10.1146/annurev-astro-081913-040025}

\bibitem[{{Sormani} {et~al.}(2017){Sormani}, {Tre{\ss}}, {Klessen}, \&
  {Glover}}]{Sormani2017}
{Sormani}, M.~C., {Tre{\ss}}, R.~G., {Klessen}, R.~S., \& {Glover}, S. C.~O.
  2017, \mnras, 466, 407, \dodoi{10.1093/mnras/stw3205}

\bibitem[{{Steigman} {et~al.}(1975){Steigman}, {Strittmatter}, \&
  {Williams}}]{Steigman75}
{Steigman}, G., {Strittmatter}, P.~A., \& {Williams}, R.~E. 1975, \apj, 198,
  575, \dodoi{10.1086/153636}

\bibitem[{{Stone} \& {Gardiner}(2009)}]{StoneGardiner09}
{Stone}, J.~M., \& {Gardiner}, T. 2009, \na, 14, 139,
  \dodoi{10.1016/j.newast.2008.06.003}

\bibitem[{{Stone} {et~al.}(2008){Stone}, {Gardiner}, {Teuben}, {Hawley}, \&
  {Simon}}]{Stone08_Athena}
{Stone}, J.~M., {Gardiner}, T.~A., {Teuben}, P., {Hawley}, J.~F., \& {Simon},
  J.~B. 2008, \apjs, 178, 137, \dodoi{10.1086/588755}

\bibitem[{{Su} {et~al.}(2018){Su}, {Hopkins}, {Hayward}, {Ma},
  {Boylan-Kolchin}, {Kasen}, {Kere{\v{s}}}, {Faucher-Gigu{\`e}re}, {Orr}, \&
  {Wheeler}}]{Su18}
{Su}, K.-Y., {Hopkins}, P.~F., {Hayward}, C.~C., {et~al.} 2018, \mnras, 480,
  1666, \dodoi{10.1093/mnras/sty1928}

\bibitem[{{Sutherland} \& {Dopita}(1993)}]{SutherlandDopita93}
{Sutherland}, R.~S., \& {Dopita}, M.~A. 1993, \apjs, 88, 253,
  \dodoi{10.1086/191823}

\bibitem[{{Tenorio-Tagle} {et~al.}(2019){Tenorio-Tagle}, {Silich},
  {Palou{\v{s}}}, {Mu{\~n}oz-Tu{\~n}{\'o}n}, \& {W{\"u}nsch}}]{TT19}
{Tenorio-Tagle}, G., {Silich}, S., {Palou{\v{s}}}, J.,
  {Mu{\~n}oz-Tu{\~n}{\'o}n}, C., \& {W{\"u}nsch}, R. 2019, \apj, 879, 58,
  \dodoi{10.3847/1538-4357/ab2455}

\bibitem[{{Thompson} \& {Krumholz}(2016)}]{ThompsonKrumholz16}
{Thompson}, T.~A., \& {Krumholz}, M.~R. 2016, \mnras, 455, 334,
  \dodoi{10.1093/mnras/stv2331}

\bibitem[{{Turner} {et~al.}(2017){Turner}, {Consiglio}, {Beck}, {Goss}, {Ho},
  {Meier}, {Silich}, \& {Zhao}}]{Turner17}
{Turner}, J.~L., {Consiglio}, S.~M., {Beck}, S.~C., {et~al.} 2017, \apj, 846,
  73, \dodoi{10.3847/1538-4357/aa8669}

\bibitem[{{Vink} {et~al.}(2001){Vink}, {de Koter}, \& {Lamers}}]{Vink01}
{Vink}, J.~S., {de Koter}, A., \& {Lamers}, H.~J.~G.~L.~M. 2001, \aap, 369,
  574, \dodoi{10.1051/0004-6361:20010127}

\bibitem[{{Vink} \& {Sander}(2021)}]{Vink21}
{Vink}, J.~S., \& {Sander}, A. A.~C. 2021, \mnras,
  \dodoi{10.1093/mnras/stab902}

\bibitem[{{Virtanen} {et~al.}(2020){Virtanen}, {Gommers}, {Oliphant},
  {Haberland}, {Reddy}, {Cournapeau}, {Burovski}, {Peterson}, {Weckesser},
  {Bright}, {van der Walt}, {Brett}, {Wilson}, {Millman}, {Mayorov}, {Nelson},
  {Jones}, {Kern}, {Larson}, {Carey}, {Polat}, {Feng}, {Moore}, {Vand erPlas},
  {Laxalde}, {Perktold}, {Cimrman}, {Henriksen}, {Quintero}, {Harris},
  {Archibald}, {Ribeiro}, {Pedregosa}, {van Mulbregt}, \& {SciPy 1. 0
  Contributors}}]{scipy}
{Virtanen}, P., {Gommers}, R., {Oliphant}, T.~E., {et~al.} 2020, Nature
  Methods, 17, 261, \dodoi{10.1038/s41592-019-0686-2}

\bibitem[{{Walch} \& {Naab}(2015)}]{Walch_Naab2015}
{Walch}, S., \& {Naab}, T. 2015, \mnras, 451, 2757,
  \dodoi{10.1093/mnras/stv1155}

\bibitem[{{Wall} {et~al.}(2020){Wall}, {Mac Low}, {McMillan}, {Klessen},
  {Portegies Zwart}, \& {Pellegrino}}]{Wall20}
{Wall}, J.~E., {Mac Low}, M.-M., {McMillan}, S. L.~W., {et~al.} 2020, \apj,
  904, 192, \dodoi{10.3847/1538-4357/abc011}

\bibitem[{{Weaver} {et~al.}(1977){Weaver}, {McCray}, {Castor}, {Shapiro}, \&
  {Moore}}]{Weaver77}
{Weaver}, R., {McCray}, R., {Castor}, J., {Shapiro}, P., \& {Moore}, R. 1977,
  \apj, 218, 377, \dodoi{10.1086/155692}

\bibitem[{{W{\"u}nsch} {et~al.}(2017){W{\"u}nsch}, {Palou{\v{s}}},
  {Tenorio-Tagle}, \& {Ehlerov{\'a}}}]{Wunsch17}
{W{\"u}nsch}, R., {Palou{\v{s}}}, J., {Tenorio-Tagle}, G., \& {Ehlerov{\'a}},
  S. 2017, \apj, 835, 60, \dodoi{10.3847/1538-4357/835/1/60}

\end{thebibliography}
\bibliographystyle{aasjournal}

\end{document}